\documentclass[intlimits,twoside,a4paper]{article}
\pdfoutput=1

\usepackage{graphicx}			
\usepackage{amsmath}			
\usepackage{amsfonts}
\usepackage{amssymb}
\usepackage{amsthm}
\usepackage{xcolor}
\usepackage{subfigure}
\usepackage[T2A]{fontenc}
\usepackage[cp1251]{inputenc}

\usepackage[eqsecnum]{cmpj2}

\newcommand{\dslash}{\not{\hbox{\kern-2pt $\partial$}}}
\newcommand{\bqa}{\begin{eqnarray}}
\newcommand{\eqa}{\end{eqnarray}}
\newcommand{\nn}{\nonumber \\}

\newcommand{\beq}{\begin{equation}}
\newcommand{\eeq}{\end{equation}}

\renewcommand{\vec}[1]{{\boldsymbol{#1}}}

\def\be{\begin{eqnarray}}
\def\ee{\end{eqnarray}}

\issue{2016}{19}{3}{33703}
\doinumber{10.5488/CMP.19.33703}

\title[Counting Majorana bound states using complex momenta]{Counting Majorana bound states using complex momenta}

\author[I. Mandal]{I. Mandal}
\address{Perimeter Institute for Theoretical Physics, 31 Caroline St. N., Waterloo ON N2L 2Y5, Canada}

\authorcopyright{I. Mandal, 2016}
\date{Received February 17, 2016, in final form April 14, 2016}

\begin{document}

\maketitle

\begin{abstract}
Recently, the connection between Majorana fermions bound to the defects in arbitrary dimensions, and complex momentum roots of the vanishing determinant of the corresponding bulk Bogoliubov–de Gennes (BdG) Hamiltonian, has been established (EPL, 2015, \textbf{110}, 67005). Based on this understanding, a formula has been proposed to count the number ($n$) of the zero energy Majorana bound states, which is related to the topological phase of the system. In this paper, we provide a proof of the counting formula and we apply this formula to a variety of 1d and 2d models belonging to the classes BDI, DIII and D. We show that we can successfully chart out the topological phase diagrams. Studying these examples also enables us to explicitly observe the correspondence between these complex momentum solutions in the Fourier space, and the localized Majorana fermion wavefunctions in the position space. Finally, we corroborate the fact that for systems with a chiral symmetry, these solutions are the so-called ``exceptional points'', where two or more eigenvalues of the complexified Hamiltonian coalesce.

\pacs 73.20.-r, 74.78.Na, 03.65.Vf

\end{abstract}

\section{Introduction}

Topological superconductors \cite{Hasan_2010} are systems which can provide the condensed matter version of Majorana fermions, because they can  host topologically protected zero energy states at a defect or edge, for which the creation operator ($\gamma^{\dagger}_{E=0}$) is equivalent to the annihilation operator ($\gamma_{E=0}$). These localized zero-energy states obey  non-Abelian braiding statistics \cite{Kitaev_2001,Kitaev_2003}, which can find potential applications in designing fault-tolerant topological quantum computers \cite{Kitaev_2001,Nayak_2008}. Although Majorana fermion bound states have not yet been conclusively found in nature, they have been theoretically shown to exist in low dimensional spinless $p$-wave superconducting systems \cite{Read_Green_2000,Kitaev_2001}, as well as other systems involving various heterostructures with proximity-induced superconductivity which are topologically similar to them~\cite{Fu_2008,Zhang_2008,Sau,Long-PRB,Roman,Oreg,abel}.

Non-interacting Hamiltonians for gapped topological insulators and topological superconductors, in arbitrary spatial dimensions, can be classified into ten topological symmetry classes \cite{ludwig,Ryu_2010,kitaev2}, characterized by certain topological invariants. Moreover, there exists a unified framework for classifying topological defects in insulators and superconductors \cite{teo}, which follows from the bulk-boundary correspondence and identification of the protected gapless fermion excitations with topological invariants characterizing the defect. Here we focus on 1d and 2d Bogoliubov-de Gennes (BdG) Hamiltonians with the particle-hole symmetry (PHS) operator squaring to $+1$, which can be categorized \cite{ludwig} into three classes: BDI, DIII and D.

In our earlier work \cite{ipsita-proof}, we have explored the connection between the complex momentum solutions of the determinant of a bulk BdG Hamiltonian ($H_{\text{{BdG}}}$) in arbitrary dimensions, and the Majorana fermion wavefunctions in the position space associated with a defect or edge. We have found that the imaginary parts of these momenta are related to the exponential decay of the wavefunctions, localized at the defects, and hence their sign-change at a topological phase transition point signals the appearance or disappearance of Majorana zero mode(s). Based on this understanding, we have proposed a formula to count the number ($n$) of the zero energy Majorana bound states, which is related to the topological phase of the system. This formula serves as an alternative to the familiar $\mathbb{Z}$ and $\mathbb{Z}_2$ topological invariants \cite{ludwig,Ryu_2010,TewariPRL2012} and other counting schemes \cite{luiz,jelena1,jelena2,jelena3}.

In this paper, we prove this formula and apply it to a variety of 1d and 2d models belonging to the classes BDI, DIII and D. We show that we can successfully chart out the topological phase diagrams. Studying these examples also enables us to explicitly observe the correspondence between these complex momentum solutions in the Fourier space, and the localized Majorana fermion wavefunctions in the position space. Finally, we also corroborate the fact that for systems with a chiral symmetry, these solutions can be identified with the so-called ``exceptional points'' (EP's)  \cite{Panch,Berry,Kato,Rotter,Heiss,Fagotti,Brand}, where two or more eigenvalues of the complexified Hamiltonian coalesce. EP's are singular points at which the norm of at least one eigenvector vanishes, when certain real parameters appearing in the Hamiltonian are continued to complex values, and the complexified Hamiltonian becomes non-diagonalizable. The concept of EP's is similar to that of a degeneracy point, but with the important difference that all the
energy eigenvectors cannot be made orthogonal to each other. In previous works, EP's have been used \cite{sourin,dmitry1,dmitry2,ramon,ipsita} to describe topological phases of matter for 1d topological superconductors/superfluids.

The paper is organized as follows: in section~\ref{formula}, we review the results obtained earlier \cite{ipsita-proof} for counting the number ($n$) of Majorana zero modes bound to defects, based on the bulk-edge correspondence. In section~\ref{proof-formula}, we provide a proof of the counting formula. In section~\ref{bdi}, we consider some 1d and 2d models in the class BDI and apply the EP formalism to count $n$. Section~\ref{d3} is devoted to the study of edge states for Hamiltonians in class DIII, where we illustrate the applicability of EP solutions as the chiral symmetry exists. In section~\ref{classd}, we discuss some systems in the class D and conclude that EP's cannot be related to the Majorana fermion wavefunctions for such Hamiltonians, because chiral symmetry is broken. We conclude with a summary and outlook in section~\ref{conclusion}. In appendix~\ref{append:eg}, we provide a simple example to show how one should choose the correct EP solutions such that their imaginary parts are continuous functions in the parameter space in order to evaluate our counting formula.

\section{Counting formula for the Majorana zero modes}
\label{formula}

In this section, we review the connection~\cite{ipsita-proof} between the complex momentum solutions of \linebreak$\det[H_{\text{BdG}}(\textbf k)]=0$,  and the Majorana fermion wavefunctions in the position space associated with a defect or edge.

We consider a topological defect embedded in (or at the boundary of) a $d$-dimensional topological superconductor. Let $m$ be the dimensions of the defect, parametrized by the Cartesian coordinates $\textbf {r}_{\perp} =\left (  r_{1}, \ldots, r_{d-m}   \right )$ and $\textbf {r}_{\parallel} =\left (  r_{d-m+1}, \ldots, r_d   \right )$, located at $\textbf r_{\perp} = 0 $. Let $\textbf {k}_{\perp} =k_{\perp} \hat {\vec \Omega} = \left (  k_{1}, \ldots, k_{d-m}   \right ) $ and $\textbf {k }_{\parallel} =\left (  k_{d-m+1}, \ldots, k_d   \right )$ be the corresponding conjugate momenta, where $k_{\perp} = |\textbf k_{\perp}|$ and $ \hat {\vec \Omega}$ is the unit vector when written in spherical coordinates.

For a generic $H_{\text{{BdG}}}$, let $ k_\text{A}^{j} $ and $k_\text{B }^{j}$  ($j = 1, \ldots , Q $) be the two sets of complex $k_{\perp}$-solutions for \linebreak$\det [H_{\text{{BdG}}} (\textbf{k}) ] =0 $, related by $ \lbrace  \Im (k_{ \text{A} }^{j} ) \rbrace = - \lbrace  \Im (k_{ \text{B} }^{j } ) \rbrace$, after $k_{\perp}$ has been analytically continued to the complex plane. One should be careful to choose solutions such that their imaginary parts are continuous functions of the parameter(s) which tune(s) through the transition, and the solutions in one set are related to the other by changing the sign of their imaginary parts throughout. This point has been illustrated by an example in appendix~\ref{append:eg}. Assuming the Majorana wavefunction to be of the form $ \sim \exp \left( - z \, |\textbf {r}_{\perp}| \right )$ in the bulk, the correspondence $\ri k_{\perp} \leftrightarrow - z$ has been established \cite{ipsita-proof}.  At a topological phase transition point, one or more of the $\Im (  k_\text{A/B}^j ) $'s go through zero. When $\Im  (  k_\text{A/B}^j ) $ changes sign at a topological phase transition point, the position space wavefunction of the corresponding Majorana fermion changes from exponentially decaying to exponentially diverging or vice versa. If the former happens, the Majorana fermion ceases to exist. A new Majorana zero mode appears in the latter case. The count ($n$) for the Majorana fermions for a defect is captured by the function
\begin{eqnarray}
\label{fchiral1}
f(\lbrace \lambda_i \rbrace, \textbf k_{\parallel} , \hat{\vec{\Omega}} ) =  \frac{1} {2} \,
\Bigg| \sum_{j=1}^{Q} \Bigg(\text{sign} \left \lbrace \Im \left [ k_\text{A/B}^{j} \big( \lbrace \lambda_i \rbrace   , \textbf k_{\parallel} , \hat{\vec{\Omega}} \big) \right ] \right \rbrace
 - \text{sign} \left \lbrace \Im \left [ k_\text{A/B}^{j} \big ( \lbrace \lambda_i^0 \rbrace   , \textbf k_{\parallel}^0 , \hat{\vec{\Omega}}^0 \big ) \right ] \right \rbrace
\Bigg) \,  \Bigg |\,,
\end{eqnarray}
where $(\lbrace \lambda_i \rbrace, \textbf k_{\parallel} , \hat{\vec{\Omega}} ) $ are the parameters appearing in the expressions for $k_\text{A/B}^{j}\,$, and $(\lbrace \lambda_i^0 \rbrace, \textbf k_{\parallel}^0 , \hat{\vec{\Omega}}^0 )$ are their values at any point in the non-topological phase.

If there is a chiral symmetry operator $\mathcal{O}$ which anticommutes with the Hamiltonian, the latter takes the form
\begin{equation}
H_{\text{{chiral}}} (\textbf k) = \left(
\begin{array}{cc}
 0 & \mathcal{A}  (\textbf k) \\
 \mathcal{A}^{\dagger} (\textbf k)  &  0   \\
\end{array} \right) ,
\label{chiralmom}
\end{equation}
in the momentum space, for the corresponding bulk system with no defect.  On analytically continuing the magnitude $k_{\perp} \equiv k= |\textbf k| $ to the complex $k_{\perp}$-plane, at least one of the eigenvectors of $H_{\text{{chiral}}}( \textbf k)$ collapses to zero norm where
\begin{equation}
\det \left[ \mathcal{A} (\textbf{k}) \right] =0 \qquad \text{or} \qquad
\det \big[  \mathcal{A}^{\dagger} (\textbf{k}) \big] =0.
\label{detak}
\end{equation}
These points are associated with the solutions of EP's for complex $k_{\perp}$-values where two or more energy levels coalesce. Furthermore, these coalescing eigenvalues have zero magnitude since $\det [ \mathcal{A} ( \textbf{k} ) ] =0$ (or $ \det [ \mathcal{A}^{\dagger} (\textbf{k})] =0$) also implies $\det[H_{\text{{chiral}}} (\textbf{k}) ] =0 $. $H_{\text{{chiral}}} (\textbf k)$ becomes non-diagonalizable, as in the complex $k_{\perp}$-plane, $\det[ \mathcal{A} (\textbf{k})] = 0 \nRightarrow  \det [ \mathcal{A}^{\dagger} (\textbf{k}) ] = 0$  (or vice versa). However, at the physical phase transition points, the imaginary parts of one or more solutions vanish, and $\det[\mathcal{A} ( \textbf{k} )] = \det [\mathcal{A}^{\dagger} (\textbf{k}) ] = 0$ for those solutions, making $H_{\text{{chiral}}} (\textbf k)$ once again diagonalizable and marking the disappearance of the corresponding EP's.

Since it satisfies equation~(\ref{detak}), each EP solution corresponds to a Majorana fermion of a definite chirality with respect to $\mathcal O$.
If $\mathcal{A}^{\dagger} (\textbf{k})= \mathcal{A}^{\text{T}} (-\textbf{k}) $ holds, then the two sets of EP's are related by $ \lbrace k_{\text{ A} }^{j} \rbrace = - \lbrace  k_{ \text{B} }^{j } \rbrace$, one set corresponding to the solutions obtained from one of the two off-diagonal blocks. In such cases, the pairs of the Majorana fermion wavefunctions are of opposite chiralities.

\section{Derivation of the counting formula}
\label{proof-formula}

A simple derivation of the counting formula in equation~(\ref{fchiral1}) can be motivated as follows:
\begin{enumerate}

\item
Let us consider one of the solutions given by $j=1$.
In the non-topological phase, say phase ``$0$'', $k_\text{A}^{1} ( \lbrace \lambda_i^0 \rbrace   , \textbf k_{\parallel}^0 , \hat{\vec{\Omega}}^0  )   $ gives no Majorana zero mode and hence does not give rise to any decaying mode localized at a defect. On the other hand, in a topological phase, say phase ``$\text{t}$'', with a Majorana  wavefunction $\sim \exp [ - |\Im (k_\text{A }^{ 1 } ) | \, r_{\perp}] $, $ k_\text{A}^{ 1 } ( \lbrace \lambda_i \rbrace   , \textbf k_{\parallel} , \hat{\vec{\Omega}} ) \big |_{\text{phase\,t}} $ localized at $ r_{\perp} =0$ and zero at $ r_{\perp} = \infty $, should now give rise to an admissible decaying zero mode solution. This implies that there is a change in sign of $ \Im ( k_\text{A}^{1} )$ from $ - 1$ to $+ 1$ when we jump from phase ``$0$'' to phase ``$\text{t}$''.

\item
Majorana zero modes must occur in pairs, though they might be localized far apart. Hence, if $  k_\text{A}^{1} \big |_{\text{phase\, C} } $ corresponds to a Majorana mode localized at $ r_{\perp}=0 $, then $ k_\text{B}^{1} \big |_{\text{phase\, t} }$ must correspond to one localized at $r_{\perp}= \infty$, where $k_\text{B}^1 = (k_\text{A}^1 )^*$. Hence, whether or not we are in the topological phase ``$\text t$'' is captured by the function $ f_1 = \frac{1} {2} \,
 \big | \text{sign} \big \lbrace \Im [ k_\text{A/B}^{1} ( \lbrace \lambda_i \rbrace   , \textbf k_{\parallel} , \hat{\vec{\Omega}} )  ] \big \rbrace
 - \text{sign} \big \lbrace \Im [ k_\text{A/B}^{ 1 }  ( \lbrace \lambda_i^0 \rbrace   , \textbf k_{\parallel}^0 , \hat{\vec{\Omega}}^0 ) ] \big \rbrace \big |$ taking the value $ 1$ or zero.

 \item
 From the above discussion, it may seem that the counting formula should be given by
\begin{eqnarray}
\frac{1} {2}\sum_{j=1}^{Q}  \Big | \text{sign} \left \lbrace \Im \left [ k_\text{A/B}^{j} \big( \lbrace \lambda_i \rbrace   , \textbf k_{\parallel} , \hat{\vec{\Omega}} \big) \right] \right \rbrace
- \text{sign} \left \lbrace \Im \left [ k_\text{A/B}^{j} \big ( \lbrace \lambda_i^0 \rbrace   , \textbf k_{\parallel}^0 , \hat{\vec{\Omega}}^0 \big ) \right] \right \rbrace
 \Big | \,. \nonumber
 \end{eqnarray}
 However, this is not quite correct. To understand this, let us consider the scenario when at least two of the solutions, say $k_\text{A}^{ 1 }$ and $ k_\text{A}^{ 2 }$ are such that $ \text{sign} [\Im ( k_\text{A}^{2})] \big |_{\text{phase}\, 0}
 = - \text{sign}[\Im ( k_\text{A}^{1}) ]\big |_{\text{phase}\, 0 } $. This implies that in the trivial phase, the wavefunction given by $ c_1\exp ( \ri \, k_\text{A}^{1}   \big |_{\text{phase}\, 0 } \, r_{\perp} ) + c_2  \, \exp ( \ri \, k_\text{A}^{2}   \big |_{\text{phase}\, 0} \, r_{\perp} )$ is inadmissible for not being capable of satisfying the boundary conditions --- the only solution is $c_1=c_2 =0 $.
In another topological phase, say ``$\, \tilde{\text t}\,$'', let $ \text{sign} [\Im ( k_\text{A}^{ 1 })]\big |_{\text{phase}\, \tilde{\text t} }
 = - \text{sign} [\Im ( k_\text{A}^{ 1 } ) ]\big |_{\text{phase}\, 0} $
and $ \text{sign} [\Im ( k_\text{A}^{ 2 }) ]\big |_{\text{phase}\, \tilde{\text t} }
 = - \text{sign}[\Im ( k_\text{A}^{ 2 })]\big |_{\text{phase}\, 0} $. This means that both
 $ \Im ( k_\text{A}^{ 1 } ) $ and $ \Im ( k_\text{A}^{ 2 })$ change sign when we jump from phase ``$0$''
 to phase ``$\, \tilde{\text t}\,$''. However, they still should not give any Majorana zero mode in the phase ``$\, \tilde{\text t}\,$'', because $ \tilde c_1 \, \exp (\ri \, k_\text{A}^{1}   \big |_{\text{phase}\, \tilde{\text t}} \, r_{\perp}) +
 \tilde c_2  \, \exp (\ri \, k_\text{A}^{2}   \big |_{\text{phase}\, \tilde{\text t}} \, r_{\perp} ) $ cannot satisfy the boundary conditions. So, the correct formula is given by equation~(\ref{fchiral1}).

 \end{enumerate}


\section{EP formalism for the BDI class}
\label{bdi}

In this section, we consider some 1d and 2d spinless models in the BDI class, which can support multiple Majorana fermions at any end of an open chain. For systems in this class, there exists a chiral symmetry operator $\mathcal{O}$, such that $H_{\text{{BdG}}}$ can be rotated to the form $H_{\text{{chiral}}}$ in equation~(\ref{chiralmom}).

After reviewing the transfer matrix scheme to find Majorana fermion solutions localized at an edge, we show how EP solutions in the complex $k_{\perp}$-plane can be used to count the number of Majorana zero modes in a given topological phase. We also make emphasis on the connection of these EP solutions with the position space wavefunctions calculated in the real space lattice with open ends.

\subsection{Transfer matrix approach}
\label{transfer}

Kitaev \cite{Kitaev_2001} suggested the model of a 1d $p$-wave superconducting
chain, which can support Majorana zero modes at the two ends. For a finite and open chain with $N$ sites, the Hamiltonian
takes the form
\begin{eqnarray}
\label{kitaev-open}
H_K  = - \sum_{j=1}^{N} \mu \left(  c_j^{\dagger}\, c_j - \frac{1}{2} \right) \
+ \sum_{j=1}^{N-1 }  \left( - w \, c_j^{\dagger}\, c_{j+1} + \Delta \, c_j \, c_{j+1} + \text{h.c.} \right) \,,
\end{eqnarray}
where $\mu$ is the chemical potential, $w$ and $ \Delta $ are the nearest-neighbour hopping amplitude and superconducting gap, respectively. The pair of fermionic annihilation and creation operators, $c_j$ and $c_{j}^{\dagger}$, describe the lattice site $j$, and obey the usual anticommutation
relations $ \lbrace c_j,c_j' \rbrace = 0$ and $ \lbrace c_j ,c_{j'}^{\dagger} \rbrace = \delta_{j j'} $.
The Majorana mode structure of the wire can be better understood by rewriting the above Hamiltonian in terms of the Majorana operators
\begin{eqnarray}
a_j = c_j^{\dagger} +  c_j \,,
\qquad b_j  = -\ri \left( c_j^{\dagger} - c_j \right)\,,
\end{eqnarray}
satisfying
\begin{eqnarray}
a_j = a_j^{\dagger} \,, \qquad b_j = b_j^{\dagger} \,,  \qquad \lbrace a_j ,b_{j'} \rbrace = 0 \,,
\qquad \lbrace a_j ,a_{j'} \rbrace = \lbrace b_j ,b_{j'} \rbrace = 2\, \delta_{j j'} \,. \nonumber
\end{eqnarray}
Then, the Hamiltonian reduces to
\begin{eqnarray}
\label{kitaev-open2}
H_K  = -\frac{\ri}{2} \sum_{j=1}^{N} \mu \, a_j \, b_j
-\frac{\ri}{2} \sum_{j=1}^{N-1} \big[
(w-\Delta )  \, a_j \, b_{j+1}
- (w +\Delta )   \, b_{j} \, a_{j+1}
\big]
 \,.
\end{eqnarray}

This chain can support one Majorana bound state (MBS) at an edge for appropriate values of the parameters.
More recently, a variation of the model was considered with next-nearest-neighbour hopping and pairing amplitudes \cite{sudip}. A general version of such longer-ranged interactions with all possible hoppings and pairings was studied \cite{diptiman1,diptiman2} with the Hamiltonian
\begin{eqnarray}
\label{diptiman-open}
H_l =  -\frac{\ri}{2} \sum_{j=1}^{N} \mu \, a_j \, b_j
- \ri\sum_{r= 1 }^{q} \sum_{j=1 }^{N-q}
\big [J_{-r} \, a_j \, b_{j+r} + J_{r} \, a_{j+r} \, b_{j}  \big ]\,,
\end{eqnarray}
where the $J_{\pm r}$'s are real parameters, and $ 0< q < N $. These models can support multiple MBSs at an edge. If we impose periodic boundary conditions (PBC's), the Hamiltonian can be diagonalized by a Bogoliubov transformation:
\begin{eqnarray}
&H_l = - \displaystyle\sum_{ k }
\left(
\begin{array}{cc}
 c_k^{\dagger}  & c_{-k}
\end{array} \right)
\, h_l (k)
\left(
\begin{array}{c}
 c_k  \\
 c_{-k}^{\dagger}
\end{array} \right) ,& \nonumber
\\ &h_l(k) = -2 \displaystyle\sum_{r= -q}^{q}
\left(
\begin{array}{cc}
 J_r \, \cos \left(  k r \right) & -\ri \, J_r \, \sin \left(  k r \right)\\
 \ri \, J_r \, \sin \left(  k r \right)  &  -J_r \, \cos \left(  k r \right)   \\
\end{array} \right) \,, \qquad
J_0  =-\frac{\mu}{2} \,,&
\label{diptiman-bdg}
\end{eqnarray}
where the anticommuting fermion operators $( c_k^{\dagger}, \, c_k)$ are suitable linear
combinations in the momentum space of the original $( c_j , \,c_{j}^{\dagger})$ fermion operators.
The energy eigenvalues are given by
\begin{equation}
E_l (k) = \pm 2 \, \sqrt{
\Big [\sum_r J_r \, \cos \left(  k r \right) \Big ]^2
+    \Big[\sum_r  J_r \, \sin \left(  k r \right) \Big ]^2
 } \,.
\end{equation}
We now review the transfer matrix approach \cite{diptiman2011,sudip,diptiman1,diptiman2} to identify the number of MBSs at each end of the chain for this model. The transfer matrix can be obtained from the Heisenberg
equations of motion for the Majorana operators in equation~(\ref{diptiman-open}):
\begin{eqnarray}
2\, \ri\, \frac{\rd  a_j} {\rd  t}
= - \ri \sum_{r=-q}^{q}  J_{-r} \, b_{j+r} \,,\qquad
2\, \ri\, \frac{\rd  b_j} {\rd  t}=
   \ri \sum_{r=-q}^{ q }  J_r \, a_{j+r}  \,.
\end{eqnarray}
Assuming the time-dependence to be of the form $a_j = A_j \, e^{ -\ri E_l t} $ and $b_j = B_j \, e^{ -\ri  E_l t} $, the $E_l=0 $ (zero energy modes) are given by the recursion relation of the
amplitudes:
\begin{eqnarray}
 &&\sum_{r=-q}^{q}  J_{-r} \, b_{j+r}  = 0 \,,\qquad
\sum_{r=-q}^{q}  J_r \, a_{j+r}  =0 \,.
\end{eqnarray}
Clearly, it will suffice to solve one set of the recursive equations to obtain the solutions for both. Assuming $ A_j = \lambda_\text{A}^j $ and $ B_j = \lambda_\text{B}^j $, we get the polynomial equations
\begin{equation}
\sum_{r=-q}^q J_{r} \,\lambda_\text{A}^{q+r} =0 \,,
\qquad \sum_{r=-q}^q J_{-r} \,\lambda_\text{B}^{q+r} =0 \,.
\label{trs-soln}
\end{equation}
An MBS can exist if we have a normalizable solution, i.e., if $ |\lambda_\text{A}| <1$ or $|\lambda_\text{B}| <1$, if the solution is to be localized at the left end. Similarly, for a  mode to be localized at the right-hand end of the chain, we must have $ |\lambda_\text{A}| > 1$ or $|\lambda_\text{B}| > 1$. Depending on the number of constraint equations (or boundary conditions on the amplitudes), one should determine the number of independent MBSs at each end of the chain.

\subsection{Relation of the EP formalism with the transfer matrix approach}
\label{epspinless}

Let us apply the EP formalism \cite{ipsita,ipsita-proof} to the Hamiltonian in equation~(\ref{diptiman-bdg}). First we rotate it to the off-diagonal form
\begin{eqnarray}
h_{l,\text{od}} (k) = U_l ^{\dagger} \, h_l(k) \, U_l
= \left(
\begin{array}{cc}
 0 & A_l(k) \\
 B_l (k)  &  0   \\
\end{array} \right) \,,
\qquad U_l = \frac{\ri}{\sqrt{2}}
\left( \begin{array}{cc}
 -1 & -1 \\
 -1   &  1   \\
\end{array} \right) \,,\qquad\qquad\quad\nonumber\\
A_l (k) =  -2 \sum_{r= -q}^{q} \big[ J_r \, \cos \left(  k r \right)+ \ri \, J_r \, \sin \left(  k r \right) \big ] \,,
\qquad
B_l (k) = -2\sum_{r= -q}^{q} \big [ J_r \, \cos \left(  k r \right) - \ri \, J_r \, \sin \left(  k r \right) \big]\,.
\end{eqnarray}
The EP's where either $A_l (k)$ or $B_l (k)$ vanishes, are given by the solutions
\begin{eqnarray}
\sum_{r=-q}^q J_r \, \tilde \lambda_{\text{A}l}^{q+r} = 0,
\qquad \text{where}\qquad  \tilde \lambda_{\text{A}l} = \exp \left( \ri k_{\text{A}l}\right),
\label{al}\\
\sum_{r=-q}^q J_{-r} \, \tilde \lambda_{\text{B} l}^{q+r} = 0,
\qquad \, \text{where}\qquad  \tilde \lambda_{\text{B}l} = \exp \left( \ri k_{\text{B}l}\right).
\label{bl}
\end{eqnarray}
Comparing equations~(\ref{trs-soln}), (\ref{al}) and (\ref{bl}), it is easy to see that the solutions for EP's in the complex $k$-plane for the PBC's correspond to the
MBS solutions for the open boundary conditions (OBC's). Since
\begin{eqnarray}
|\tilde \lambda_{\text{A} l/ \text{B}l }| < 1  \qquad \Rightarrow \qquad \Im \left( k_{\text{A}l/\text{B}l}\right) > 0 \qquad \Leftrightarrow \qquad |\lambda_\text{A / B }| < 1	\,,\\
|\tilde \lambda_{\text{A} l/ \text{B}l }| > 1  \qquad \Rightarrow \qquad \Im \left( k_{\text{A}l/\text{B}l}\right) < 0 \qquad \Leftrightarrow \qquad |\lambda_\text{A / B }| > 1\,,
\label{expdecay}
\end{eqnarray}
a sign change of $\Im \left( k_{\text{A}l/\text{B}l}\right)$ indicates a topological phase transition, by which we move from a phase where an MBS can exist to the one where that particular zero mode gets destroyed. This is related to the fact that $\Im  \left(  k_{\text{A}l/\text{B}l} \right )$'s are related to the exponential decay of the MBS position space wavefunctions localized at one end of the open chain.

\begin{figure}[!t]
\begin{center}
\includegraphics[width=0.45\textwidth]{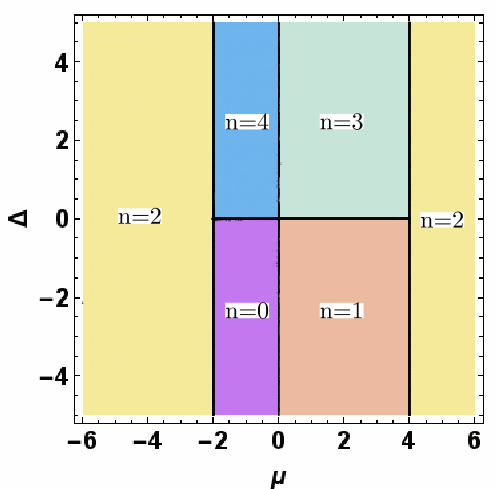}
\end{center}
\caption{\label{dip4}  (Color online)
The topological phase diagram of the Hamiltonian described by equation~(\ref{diptiman-open}), with $
J_0 = -\frac{\mu}{2} , \, J_{1}=J_{2} = \frac{ 1 + \Delta }{2}
, \, J_{- 1} = J_{- 2}=\frac{1 - \Delta} {2} $, and all other $J_r$'s  set to zero. Here, $n$ labels the number of Majorana zero modes at each end of the chain, as captured by the function $f(\mu,\Delta)$ defined in equation~(\ref{fchiral1}).
}
\end{figure}
Choosing
$
J_0 = -\frac{\mu}{2} , \, J_{1}=J_{2} = \frac{ 1 + \Delta }{2}
, \, J_{- 1} = J_{- 2}=\frac{1 - \Delta} {2} $ and all other $J_r$'s to be zero, we can get a system supporting up to four Majorana zero modes at each end of the chain. The phase diagram obtained using equation~(\ref{fchiral1}) is shown in figure~\ref{dip4}.

\begin{figure}[!b]
\begin{center}
\subfigure[]{\includegraphics[width=0.4\textwidth]{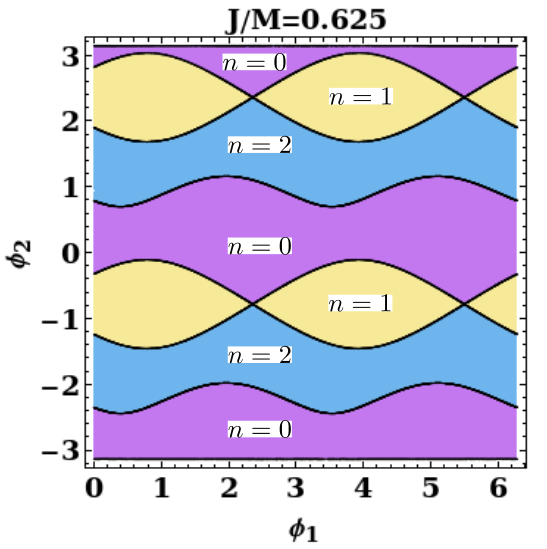}
\label{3root1}} \qquad
\subfigure[]{\includegraphics[width=0.4\textwidth]{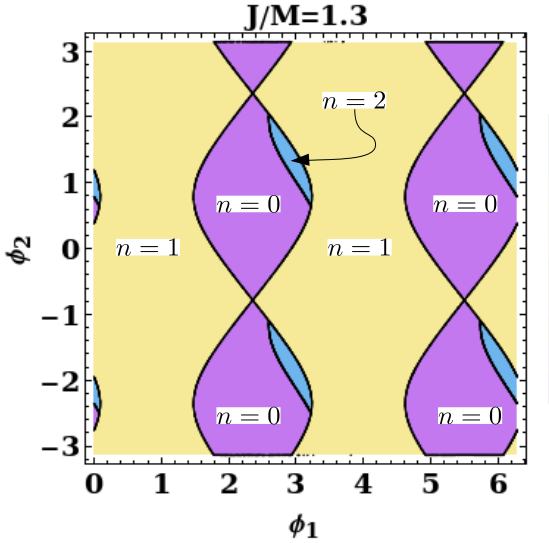}
\label{3root3}}
\end{center}
\vspace{-3mm}
\caption{\label{3root}  (Color online)
Panels (a) and (b) show the topological phase diagram of the Hamiltonian described by equation~(\ref{diptiman-open}), with $
J_0  = - \frac{ M} {2} \cos (\phi_2) , \, J_{1} = - \frac{ J } {2} \cos (\phi_1)
, \, J_{- 1} =  - \frac{ J } {2} \sin (\phi_1) $, and all other $J_r$'s  set to zero. Here, $n$ labels the number of Majorana zero modes at each end of the chain, as captured by the function $f(\mu,\Delta)$ defined in equation~(\ref{fchiral1}).
}
\end{figure}

Instead, for the parameters $J_0  = - \frac{ M}{2} \cos (\phi_2) , \, J_{1} = - \frac{ J } {2} \cos (\phi_1)
, \, J_{- 1} =  - \frac{ J } {2} \sin (\phi_1) $, and all other $J_r$'s  set to zero, we get a system having three EP's for either $A_l (k)=0$ or $B_l (k)=0$. For this model, up to two Majorana zero modes can appear at an edge. The phase diagrams for $J/M =0.625$ and $J/M=1.3$, obtained using equation~(\ref{fchiral1}), are shown in figure~\ref{3root}.

We should note another important point: if there are $Q$ EP solutions for either $A_l(k)=0$ or $B_l(k)=0$, clearly there are $2 \, Q$ solutions in total. However, for counting the zero modes in equation~(\ref{fchiral1}), we should consider only one set, where the two sets obey the relation
\begin{equation}
\label{inverse}
\tilde{\lambda}_{\text{A}l} = 1/\tilde{\lambda}_{\text{B}l}\qquad  \text{or} \qquad  k_{\text{A}l}=-k_{\text{B}l} \,.
\end{equation}
As we have already seen, these two sets correspond to the wavefunctions of the MBSs at the two opposite ends. Evidently, the MBSs exist in pairs at the two ends and the topological phase is characterized by their number at each individual end.

\subsection{Single-channel ferromagnetic nanowire}
\label{nano1}

The 1d Hamiltonian for a ferromagnetic nanowire embedded on Pb superconductor \cite{sumanta-chiral} with a single spatial channel (i.e., no transverse hopping) is given by
\begin{eqnarray}
H_{N1} = \sum_{k} \Psi_k^{\dagger}\, h_{N1} (k)\, \Psi_k \,,
\qquad   \Psi_{k} = (c_{k\uparrow},c_{k\downarrow},c_{-k\downarrow}^{\dagger},-c_{-k\uparrow}^{\dagger})^{\text T} , \qquad\qquad\qquad \nn
h_{N1} (k)   =  \xi(k) \, \sigma_0  \tau_z +  \left[ \Delta_s \, \sigma_0 + \Delta_p \sin(k)\, \textbf{d} \cdot \vec{\sigma} \right] \tau_x
+ \, \textbf{V} \cdot \vec{\sigma} \, \tau_0 \,,\qquad
\xi(k) =	 -2t \cos(k) -\mu \,.
\label{hn}
\end{eqnarray}
Here, $k$ is the 1d crystal momentum, $\Psi_{k}$ is the four-component Nambu spinor defined in the particle-hole~$(\vec{\tau})$ and spin $(\vec {\sigma} )$ spaces, and $\textbf{V}$ is the Zeeman field which can be induced by ferromagnetism. Also, $\Delta_s$ and $ \Delta_p $ are proximity-induced $s$-wave and $p$-wave superconducting pairing potentials, respectively, with $\textbf{d}$ determining the relative magnitudes of the components of the $p$-wave superconducting order parameter $\Delta_{\alpha \beta}$ $(\alpha,\beta = \uparrow , \downarrow )$.
In our calculations, we use $\textbf{d}=(1,0,0)$ and $ \textbf{V}=(0,0,V)$. This Hamiltonian belongs to the BDI class with the chiral symmetry operator given by $ \mathcal{O} = \sigma_x  \tau_y\,$.

The eigenvalues of the Hamiltonian are given by:
\begin{eqnarray}
E_1(k) = \pm \sqrt{ \xi^2(k)+V^2+ \Delta_s^2 + \Delta_p^2 \sin^2(k)  - \tilde e_1 }\,,
\qquad
E_2(k) =  \pm \sqrt{ \xi^2(k)+V^2+ \Delta_s^2 + \Delta_p^2 \sin^2(k)  + \tilde e_1 }\,,\nn
\tilde e_1 = 2\sqrt{
V^2 \left[ \Delta_s^2 + \xi^2(k) \right] + \Delta_s^2 \, \Delta_p^2 \sin^2(k)
}\,.\qquad\qquad\qquad\qquad\qquad\qquad
\label{eigennano}
\end{eqnarray}
A level crossing can occur if either $E_1 (k) =0 $ or $ E_2 (k)= 0 $. However, for a finite $V$ and $ \Delta_s$, the latter is impossible. Hence, a level crossing takes place when $E_1(k)=0$ for $ k=0 \, \, \mbox{or} \, \, \pi $ for the appropriate values of the parameters, which also indicates that this corresponds to the appearance of zero energy modes. A generic complex value of $k$ corresponding to $E_1(k)=0$ can be obtained by solving
\begin{eqnarray}
 \left [ \tilde V^2 -\xi^2(k)  + \Delta_p^2 \sin^2(k) \right ]^ 2+ 4 \, \xi^2 (k) \, \Delta_p^2 \sin^2(k) =0\,,
\qquad \tilde V = \sqrt {V^2 -\Delta_s^2 }  \,.
\label{e1sq}
\end{eqnarray}

We can rotate the Hamiltonian in equation~(\ref{hn}) to the chiral basis, where it takes the form
\begin{eqnarray}
\label{hnchiral}
H_{N1}^{\text{chi}}(k) = U_2^{\dagger} \, h_{N1}(k) \, U_2
= \left(
\begin{array}{cc}
 0 & h_u^{N1}(k) \\
 h_l^{N1} (k)  &  0   \\
\end{array} \right) \,,
\qquad U_{2} =\frac{1}{2}
\left( \begin{array}{cccc}
 -1-\ri & 0 & 1 + \ri & 0 \\
 0  & 1 + \ri & 0 & -1-\ri \\
 0 & 1-\ri & 0 & 1-\ri \\
 1-\ri & 0 & 1-\ri & 0 \\
\end{array} \right) \,,\nn \nn
h_u^{N1} (k)  =
\left(
\begin{array}{cc}
-\xi(k)+ \ri \, \Delta_p \sin(k) -V & \Delta_s \\
 \Delta_s   &   \xi(k)- \ri \,  \Delta_p \sin(k) -V   \\
\end{array} \right) \,,\qquad\qquad\qquad\qquad\nn\nn
h_l^{N1} (k)  =
\left(
\begin{array}{cc}
-\xi(k) - \ri \, \Delta_p \sin(k)-V  & \Delta_s \\
 \Delta_s  &   \xi(k)+ \ri \,  \Delta_p \sin(k) -V   \\
\end{array} \right) \,.\qquad\qquad\qquad\qquad
\end{eqnarray}
If either $\det[h_u^{N1} (k)]=0$ or $\det[h_l^{N1}(k)]=0$ for a complex $k$-value, this leads to the vanishing of the norm of one of the four eigenvectors of $H_{N1}^{\text{chi}}(k)$, signalling the existence of an EP for that value of $k$. At an EP, $H_{N1}^{\text{chi}}(k)$ is thus non-diagonalizable.

The solutions for the EP's are given by either
\begin{eqnarray}
\det\left[h_u^{N1} (k)\right]=0 \qquad
\Rightarrow \qquad
\tilde V^2 -\xi^2(k)  + \Delta_p^2 \sin^2(k) = -2\, \ri \, \xi (k) \,  \Delta_p \sin (k)\quad\nn
\Rightarrow\qquad k= k^u_{s_1,s_2} = - \ri \ln \left \{
\left[s_1 \, \tilde V -\mu + s_2 \,
\sqrt{ \left( s_1 \, \tilde V - \mu \right)^2  - 4 t^2 }\right]
\left( 2 t + \Delta_p\right)^{-1}
 \right \}\,,
 \end{eqnarray}
or
\begin{eqnarray}
\det\left[h_l^{N1} (k)\right]=0 \qquad
\Rightarrow \qquad
\tilde V^2 -\xi^2(k)  + \Delta_p^2 \sin^2(k) = 2 \, \ri \, \xi (k) \,   \Delta_p \sin (k) \qquad\nn
\Rightarrow\qquad k= k^l_{s_1,s_2} = - \ri \ln \left \{
\left[s_1 \, \tilde V -\mu + s_2 \,
\sqrt{ \left( s_1 \, \tilde V - \mu \right)^2  - 4 t^2 }\right]
\left( 2 t - \Delta_p\right)^{-1}
 \right \}\,,
 \label{hnl}
\end{eqnarray}
where $(s_1 =\pm 1$, $s_2  = \pm 1) $.
Clearly, $k=k^{u/l}_{s_1,s_2}$ also solves equation~(\ref{e1sq}), which corresponds to two coinciding zero energy solutions (where two levels coalesce \cite{Heiss} for a complex $k$-value).

\begin{figure*}[!t]
\begin{center}
\subfigure[]{\includegraphics[width=0.495\textwidth]{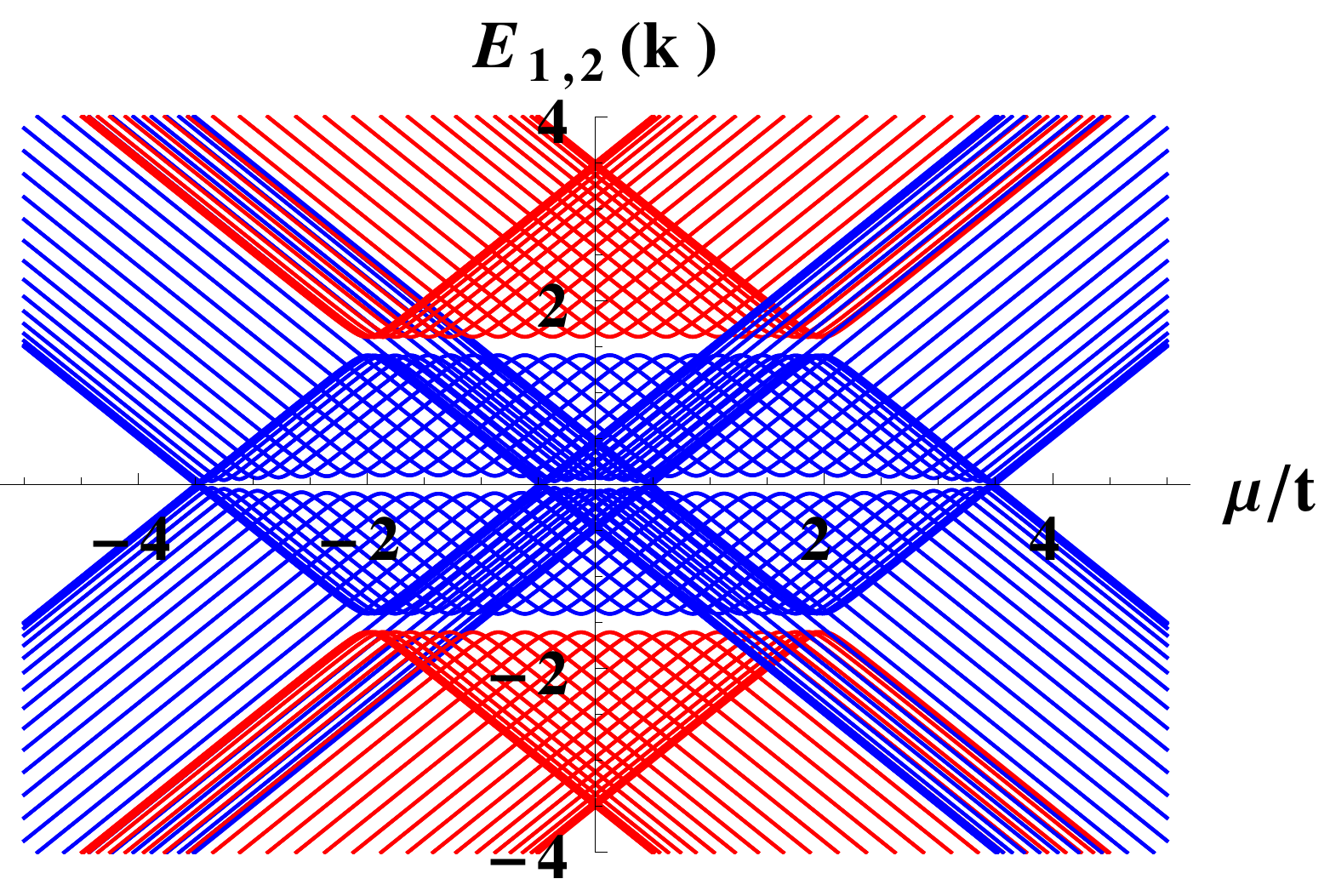}
\label{en}} \\
\subfigure[]{\includegraphics[width=0.56 \textwidth]{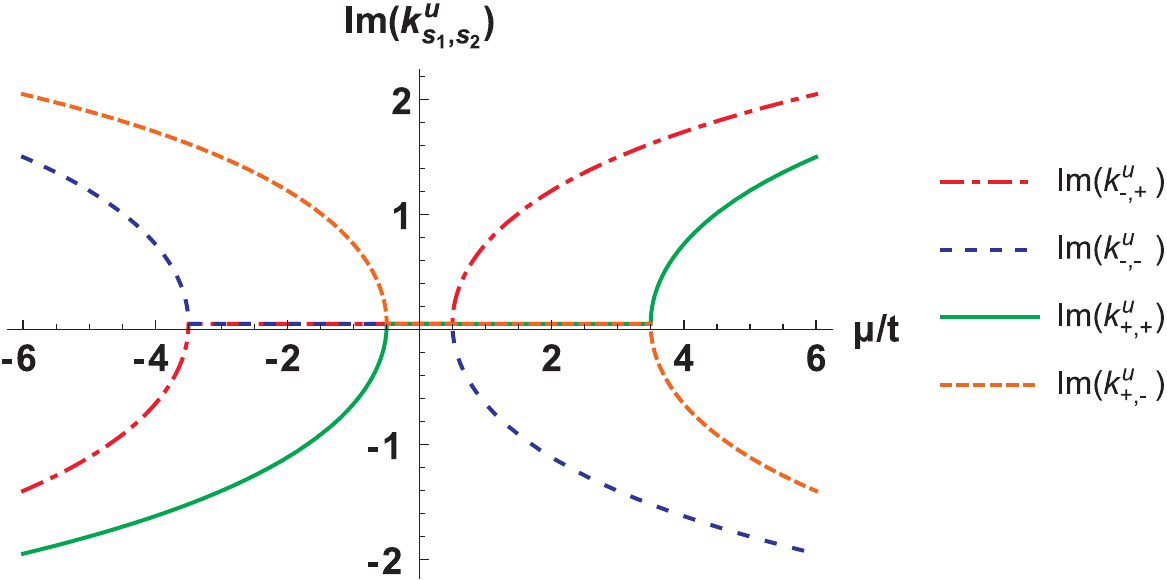}
\label{kn}} \quad
\subfigure[]{\includegraphics[width=0.4\textwidth]{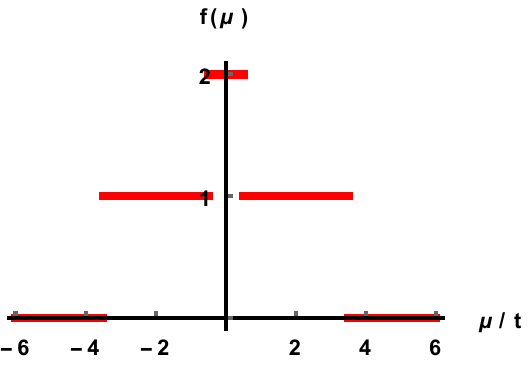}
\label{fn}}
\end{center}
\caption{\label{nano}  (Color online)
Parameters: $ \,\Delta_s = \Delta_p= 0.1 \, t$, $\tilde V= 1.5 \,t $ corresponding to the Hamiltonian in equation~(\ref{hn}).
(a) Energy bands $E_{1,2} (k)$, given in equation~(\ref{eigennano}), have been plotted in blue and red, respectively, as functions of $\mu/t$.
(b) Plots of  $ \Im \,( k^{u}_{s_1,s_2})$ versus $\mu/t$.
(c) $f(\mu)$ giving the count of the chiral Majorana zero modes as a function of $\mu/t$.
}
\end{figure*}
The plots of the energy bands, $\Im\,( k^{u}_{s_1,s_2})$, and $f(\mu)$ have been shown in figure~\ref{nano} , using the values $\Delta_s = \Delta_p = 0.1 t$ and $\tilde V = 1.5 t $.

Now, let us try to understand the existence of the EP's throughout a given topological phase and their disappearance right at the phase transition points, the latter being tied to the sign change of the  $\Im \,( k^{u/l}_{s_1,s_2} )$'s. The Hamiltonian in equation~(\ref{hnchiral}), when written in position space, gives the following equations for the Majorana zero modes, $\psi_{+} =(u_+,0)^\text{T}$ and $\psi_{-} =(0,u_-)^\text{T}$ (with chirality $+1$ and $-1$, respectively):
\begin{eqnarray}
\left(
\begin{array}{cc}
\partial_x^2 + \mu + \Delta_p \, \partial_x -V  & \Delta_s \\
 \Delta_s  &  -\partial_x^2-\mu - \Delta_p \, \partial_x -V   \\
\end{array} \right) u_+ = 0 \,,
\nn\nn
\left(
\begin{array}{cc}
\partial_x^2 + \mu - \Delta_p \, \partial_x -V  & \Delta_s \\
 \Delta_s  &  -\partial_x^2-\mu + \Delta_p \, \partial_x -V   \\
\end{array} \right) u_- = 0 \,.
\end{eqnarray}
Here, we have assumed a continuum for an open wire and set $t=1$. For $\psi_{-}\,$, let us assume the trial solution $u_- = \displaystyle\sum_{r} \exp \left( -z_r  \, x \right )
\left(
\begin{array}{cc}
u_r^{\uparrow}  \\
u_r^{\downarrow}    \\
\end{array}
\right) $. The complex $z_r$'s must satisfy the quartic equation
\begin{eqnarray}
\det \left(
\begin{array}{cc}
z_r^2 + \mu  - \Delta_p \, z_r -V  & \Delta_s \\
 \Delta_s  &  -z_r^2-\mu  +  \Delta_p \, z_r -V    \\
\end{array} \right)  = 0
\qquad \Rightarrow\qquad \left(z_r^2 + \mu  -  \Delta_p \,  z_r \right)^2 = \tilde{V}^2 \,,
\end{eqnarray}
whereas for small $k$, from equation~(\ref{hnl}), we get
\begin{equation}
\left(-k^2 + \mu + \ri \, \Delta_p \,k \right)^2 = \tilde{V}^2 \,,
\end{equation}
indicating  correspondence $ \ri k \leftrightarrow - z_r$.
The magnitude of $z_r$ will determine the admissible MBS solutions subject to OBC's (as analyzed in an earlier work \cite{dassarma}), just as in the transfer matrix analysis for the 1d~spinless lattice case. Hence, here also we have been able to establish the relation between the existence of EP's in the complex $k$-plane (for the periodic Hamiltonian) and the localized Majorana zero modes at the ends of an open chain.

\subsection{Two-channel time-reversal-symmetric nanowire system}
\label{nano2}


MBSs in a two-channel TRS nanowire proximity-coupled to an $s$-wave superconductor have been recently studied \cite{flensberg}. The low-energy model for the lowest bands of the system is described by the effective 1d $4 \times 4$ BdG Hamiltonian:
\begin{eqnarray}
\label{hn2}
H_{N2 } = \sum_{k} \Psi_k^{\dagger}\, h_{N2 } (k)\, \Psi_k \,,
\qquad  \Psi_{k} = (c_{k\uparrow},c_{k\downarrow},c_{-k\downarrow}^{\dagger},-c_{-k\uparrow}^{\dagger})^{\text{T}} ,  \qquad\quad\nn
h_{N2} = \tilde{\xi} (k) \, \sigma_0 \tau_z
+ v \left( k\, \sigma_z - p_c \, \sigma_0 \right) \tau_x
+ \textbf{B} \cdot \vec{ \sigma} \, \tau_0 \,,
\qquad \tilde{\xi} (k) = \frac{k^2} {2m} -\tilde \mu \,,
\end{eqnarray}
where $p_c$ is the momentum when the gap closes, $\textbf{B}$ is a magnetic field for the Zeeman term, and $( v, \, \tilde{\mu})$ are effective parameters. We have set $p_c = 2 v m $ for our calculations. Since this nanowire system belongs to the BDI class when $\textbf{B}$ is perpendicular to the spin-orbit-coupling direction, we will take $\textbf B= (B,0,0)$ in our analysis. Then, the chiral symmetry operator is given by $ \mathcal{O}= \sigma_z  \tau_y \,$.

The eigenvalues of the Hamiltonian are given by:
\begin{eqnarray}
E_1(k) = \pm \sqrt{ B^2 + k^2 \ v^2 + 4 \, m^2 \, v^4
+ \tilde{\xi}^2 (k)   - \tilde e_2 }\,,\qquad
E_2(k) =  \pm \sqrt{ B^2 + k^2 \, v^2 + 4 \, m^2 \, v^4
+ \tilde{\xi}^2 (k)  + \tilde e_2 }\,,\nn
\tilde e_2 = 2 \sqrt{
4 \, m^2 \, v^4 \left( B^2 + k^2 \, v^2 \right) + B^2  \, \tilde{\xi}^2 (k)
}\,.\qquad\qquad\qquad\qquad\qquad\qquad\quad
\label{eigennano2}
\end{eqnarray}
We can have two levels coalescing if $E_1 (k) =0 $ for a complex $k$-value obtained by solving
\begin{eqnarray}
\left [ B^2 + k^2 \, v^2 -\tilde \xi^2(k)  -4 \, m^2 \, v^2 \right ]^ 2 + 4 \, \tilde \xi^2 (k) \, k^2 \, v^2 =0\,.
\label{e1sq2}
\end{eqnarray}

\begin{figure}[!t]
\begin{center}
\subfigure[]{\includegraphics[width=0.45\textwidth]{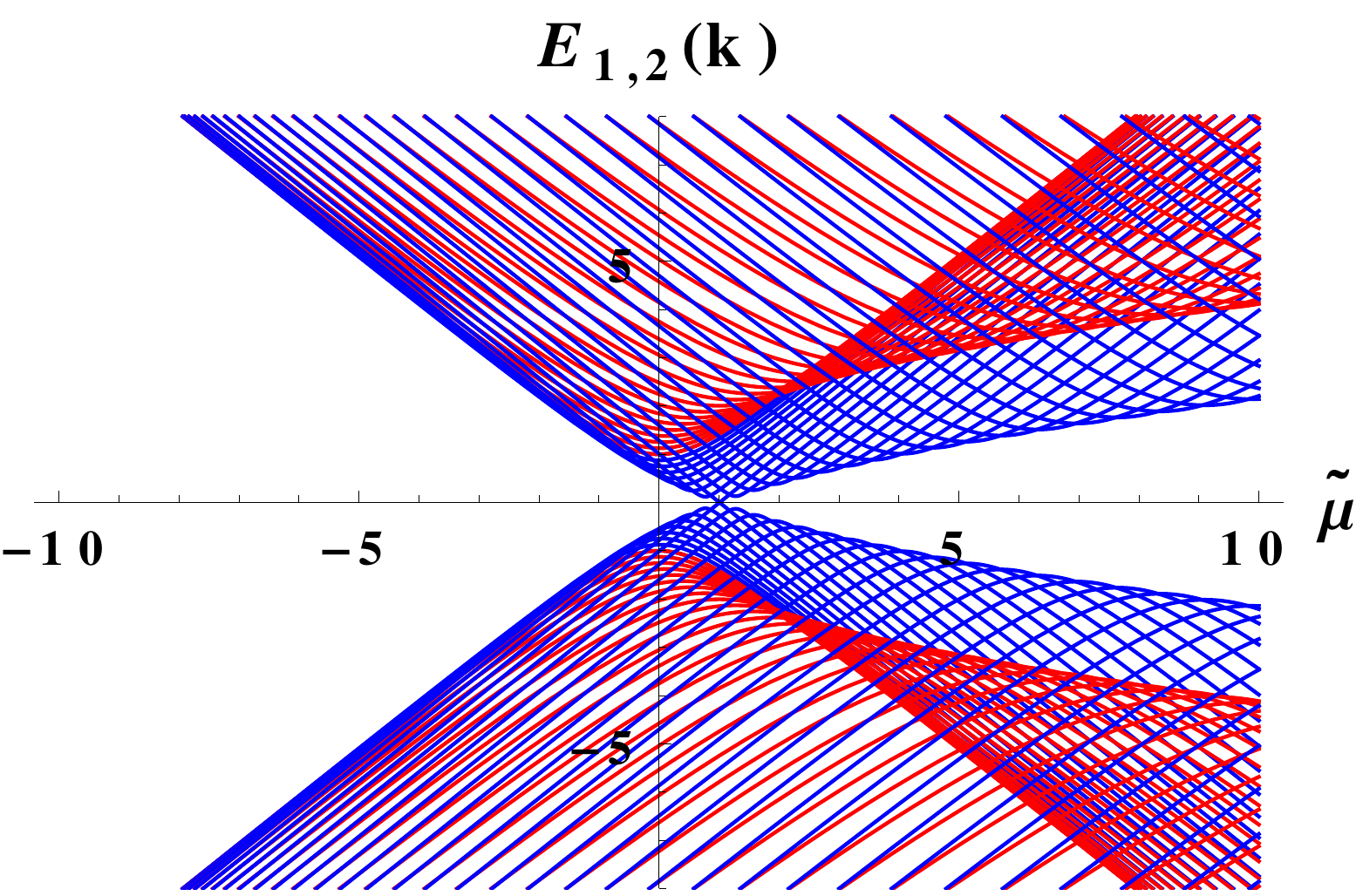}
\label{n21}} \qquad
\subfigure[]{\includegraphics[width=0.45\textwidth]{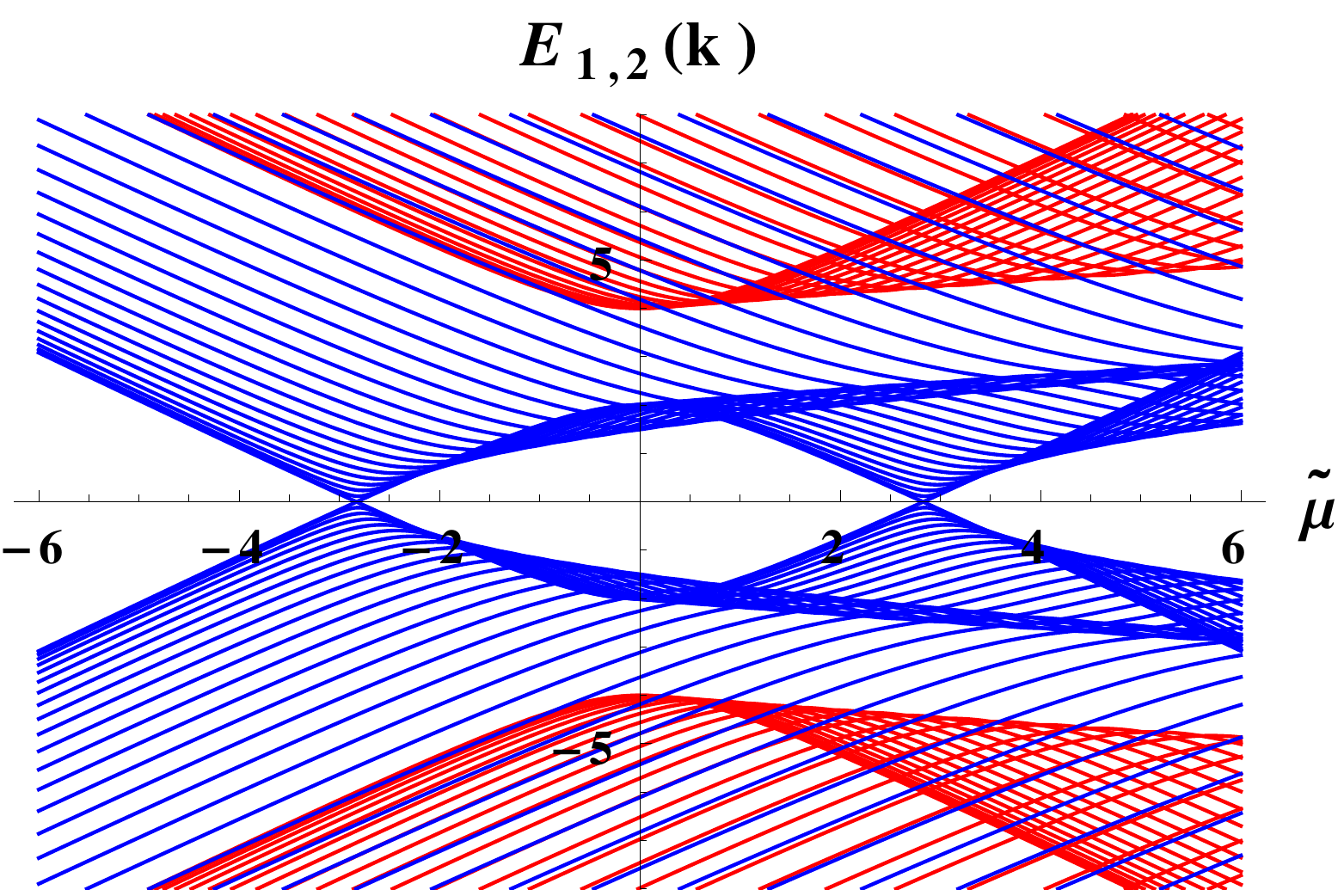}
\label{n22}}
\subfigure[]{\includegraphics[width=0.4\textwidth]{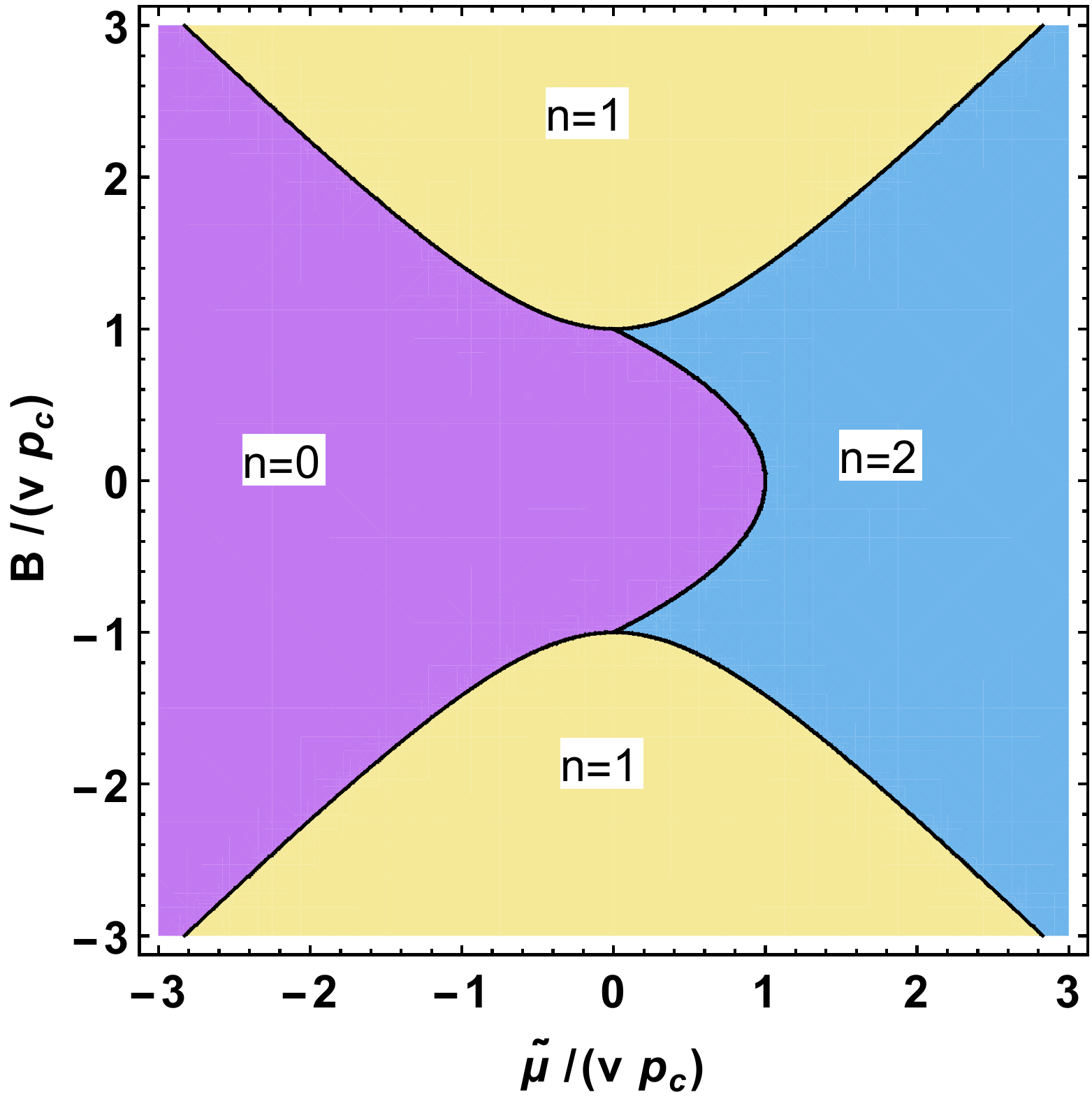}
\label{n23}}
\end{center}
\caption{\label{n2}  (Color online)
Parameters: $v=1$, $m=1/(2 v^2)$, $p_c= 2 v m$ corresponding to the Hamiltonian in equation~(\ref{hn2}).
Panels (a) and (b) show the energy bands $E_{1,2} (k)$, given in equation~(\ref{eigennano2}), as functions of $\tilde \mu$, for $B=0$ and $B=3$, respectively. $E_{1,2} (k)$ have been plotted in blue and red, respectively. Panel (c) shows the contourplot of $f(\mu)$ giving the count ``$n$'' of the MBSs in the  $\tilde{\mu}/ (v\, p_c)-B/( v\, p_c )$ plane.
}
\end{figure}

As before, we rotate the Hamiltonian in equation~(\ref{hn2}) to the chiral basis, where it takes the form
\begin{eqnarray}
\label{hnchiral2}
H_{N2}^{\text{chi}}(k) = U_3^{\dagger} \, h_{N2} (k) \, U_3
= \left(
\begin{array}{cc}
 0 & h_u^{N2}(k) \\
 h_l^{N2} (k)  &  0   \\
\end{array} \right) \,,
\qquad U_{3} =\frac{1}{2}
\left( \begin{array}{cccc}
 0 & -1-\ri &  0 & 1 + \ri \\
 0  & 1 - \ri & 0 & 1-\ri \\
 1+\ri & 0 & -1-\ri & 0 \\
 1-\ri & 0 & 1-\ri & 0 \\
\end{array} \right) \,,\nn\nn
 h_u^{N2} (k)
=
\left(
\begin{array}{cc}
-\tilde{\xi} (k)   + \ri \left( k  v +2 m v^2 \right ) & B \\
 B   &  -\tilde{\xi} (k)  + \ri \left( k  v - 2 m v^2 \right )   \\
\end{array} \right) \,,\qquad\qquad\qquad\qquad\quad\nn\nn
 h_l^{N2} (k)
= \left(
\begin{array}{cc}
-\tilde{\xi} (k)   - \ri \left( k  v +2 m v^2 \right ) & B \\
 B   &  -\tilde{\xi} (k)  - \ri \left( k  v - 2 m v^2 \right )   \\
\end{array} \right) \,.\qquad\qquad\qquad\qquad\quad
\end{eqnarray}
The solutions for the EP's are then given by either
\begin{eqnarray}
\det\left[h_u^{N2} (k)\right]=0 \qquad \Rightarrow\qquad
\left [ \tilde{\xi} (k) -\ri k  v \right ]^2 =  B^2 -4 m^2 v^4 \quad\qquad\qquad\qquad\qquad\nn
\Rightarrow \qquad k= k^u_{s_1,s_2}
= - \ri \ln \left (
 s_1
\sqrt{ m^2 v^2 -2 m \tilde \mu  + 2 \ri s_2 m \,
\sqrt{4 m^2 v^4 -B^2 }
}
+ \, \ri  m v \right )\,,
\end{eqnarray}
or
\begin{eqnarray}
\det\left[h_l^{N2} (k)\right]=0 \qquad \Rightarrow \qquad
\left [ \tilde{\xi} (k) +  \ri k v \right ]^2 =   B^2 -4 m^2 v^4 \quad\qquad\qquad\qquad\qquad\nn
\Rightarrow \qquad k= k^l_{s_1,s_2}
= - \ri \ln \left(
s_1
\sqrt{ m^2 v^2 -2 m \tilde \mu  + 2 \ri s_2 m \,
\sqrt{4 m^2 v^4 -B^2 }}
- \ri  m v   \right)\,, \label{hnl2}
\end{eqnarray}
where $ \left (s_1 =\pm 1 , s_2  = \pm 1 \right ) $.
Clearly, $k=k^{u/l}_{s_1,s_2}$ also solves equation~(\ref{e1sq2}) and hence corresponds to the coalescing of two energy levels at the zero value in the complex $k$-plane. Choosing $v=1$ and $m=1 /(2 \, v^2)$, the energy bands for $B=0$ and $B=3$, and the contourplot for $f(\tilde \mu, B)$ [defined in equation~(\ref{fchiral1})] have been shown in figure~\ref{n2}. Once again we find that $f(\tilde \mu, B)$ gives the correct topological phase diagram in figure~\ref{n2}~(c). Needless to add that here also $ \exp ( \ri\, k^{u/l}_{s_1,s_2} )$'s determine the admissible solutions for the MBS wavefunctions in the position space, at the ends of an open chain.

\subsection{Majorana edge modes for the Kitaev honeycomb model}
\label{honeycomb}

In this subsection, we consider the EP-formalism for a 2d~lattice Hamiltonian in the class BDI. The Kitaev honeycomb model \cite{kitaev2006} can be mapped onto free spinless fermions with $p$-wave pairing on a honeycomb lattice, using the Jordan-Wigner transformation. The solutions for the edge modes for a semi-infinite lattice\footnote{We would like to point out that this system is different from two-dimensional $p_x + \ri p_y$ fermionic superfluids, whose excitation spectra include gapless Majorana-Weyl fermions \cite{volovik}. Volovik showed that in such chiral superfluids, the fermionic zero modes along the domain wall have the same origin as the fermion zero modes appearing in the spectrum of the Caroli-de Gennes-Matricon bound states in a vortex core \cite{jackiw,galitski,gurarie,rahul,tewari-index}. This correspondence can be understood by picturing the chiral fermions as orbiting around the vortex axis, analogous to the motion along a closed domain boundary.} have been studied earlier \cite{nakada1996,Kohmoto2007,cano2013,krishnendu}. The momentum space Hamiltonian in terms of the Majorana operators is
\begin{eqnarray}
H_h  = \sum_{ \textbf{k}}
~\left( \begin{array}{cc}
\hat a_{\textbf{k}}^{\dagger} & \hat b_{\textbf{k}}^{\dagger}
\end{array} \right) ~h_h (\textbf k) ~
\left(
\begin{array}{c}
\hat a_{\textbf{k}} \\
\hat b_{\textbf{k}}
\end{array} \right),
\qquad
h_{h} (\textbf{k}) = \left(
\begin{array}{cc}
0 & A (\textbf k )   \\
B (\textbf k )  & 0 \\
\end{array} \right) ,
\qquad  \textbf{k} = (k_x, k_y) \,, \quad\qquad\nn
A (\textbf k ) = -2 \ri \,\left [\,J_3+J_1 \cos  \left( \frac{k_x-k_y} {2} \right)
+ J_2 \cos \left( \frac{k_x + k_y} {2} \right)  \right ]
+ 2 \,  \left[
\,J_1 \sin \left( \frac{k_x-k_y} {2} \right) + J_2 \sin \left( \frac{k_x + k_y} {2} \right) \right ]\,,\nn
B (\textbf k ) = 2 \ri \,\left [\,J_3+J_1 \cos  \left( \frac{k_x-k_y} {2} \right)
+ J_2 \cos \left( \frac{k_x + k_y} {2} \right)  \right ]
+ 2 \,  \left[
\,J_1 \sin \left( \frac{k_x-k_y} {2} \right) + J_2 \sin \left( \frac{k_x + k_y} {2} \right) \right ]\;\;\;
\label{hmom}
\end{eqnarray}
with the eigenvalues
\begin{eqnarray}
E (\textbf{k}) = \pm \,2 \,
\left\{\left[ J_3 + J_1 \cos \left( \frac{k_x-k_y} {2} \right)
+  J_2 \cos \left( \frac{k_x + k_y} {2} \right)  \right]^2
+ \, \left[ J_1 \sin \left( \frac{k_x-k_y} {2} \right) + J_2 \sin \left( \frac{k_x + k_y} {2} \right)  \right]^2    \right\}^{1/2}\,.
\label{disp1}
\end{eqnarray}

\begin{figure}[!b]
\begin{center} \quad
\subfigure[]{\includegraphics[width=0.36\textwidth]{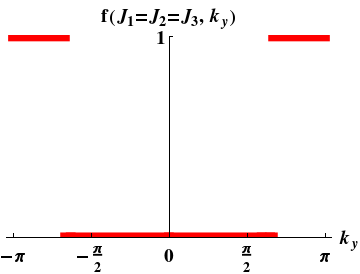}
\label{zig1}} \qquad
\subfigure[]{\includegraphics[width=0.56\textwidth]{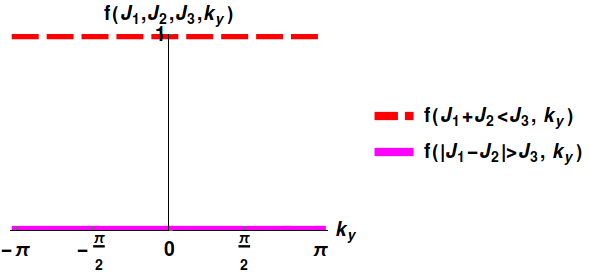}
\label{zig2}} \\
\subfigure[]{\includegraphics[width=0.36\textwidth]{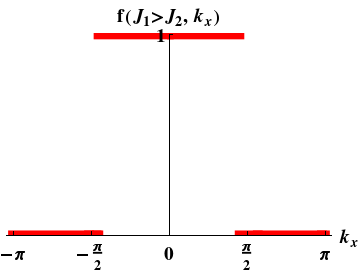}
\label{arm1}} \qquad
\subfigure[]{\includegraphics[width=0.36\textwidth]{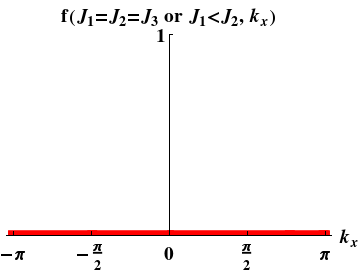}
\label{arm2}}
\end{center}
\caption{\label{zigarm}  (Color online)
Topological phase diagrams for edges for the 2d honeycomb lattice described by equation~(\ref{hmom}), as captured by the function $f(J_1,J_2,J_3,k_{\parallel})$ defined in equation~(\ref{fchiral1}). Panels (a) and (b) show the number of chiral Majorana zero modes for a zigzag edge, while panels (c) and (d) show the same for an armchair edge located at the top of a semi-infinite lattice.
}
\end{figure}

We will consider two kinds of edges \citep{nakada1996,Kohmoto2007}, namely, zigzag and armchair, which can support Majorana fermions. We will find the phase diagram using the EP's corresponding to these edges setting either $A (\textbf k ) =0$ or $B (\textbf k ) =0$, after complexifying the momentum component perpendicular to the edge. For this 2d case, $f$ in equation~(\ref{fchiral1}) is a function of $(J_1,J_2 , J_3, k_{\parallel})$, where $k_{\parallel}$ is the momentum along the 1d edge being considered.

One can have a zigzag edge in the $y$-direction, according to the convention of Nakada et al. \cite{nakada1996}, so that we will complexify $k_{\perp} = k_x$, and $ k_{\parallel} = k_y$ will be one of the parameters determining the topological phase transition points. The solution for $B (k_x=k_{\perp}, k_y=k_{\parallel} ) =0$ is given by
\begin{equation}
k_{\perp} = -2 \, \ri \ln \left [ -\frac{J_1 \,\exp (\ri \, k_{\parallel}/ 2) + J_2 \, \exp (-\ri\, k_{\parallel} /2)} {J_3} \right ] \,.
\label{zigzagep}
\end{equation}
Majorana zero modes exist for all the values of $k_y$ if $ J_1 + J_2< J_3 $. There is no edge state if $ |J_1 - J_2 | > J_3 $. For $ J_1 = J_2 = J_3 $ , edge states exist if $|k_y | > 2 \pi /3 $. These results have been plotted in figures~\ref{zigarm}~(a) and \ref{zigarm}~(b).

For the armchair edge \cite{nakada1996} in the $x$-direction on the top of the lattice, we will complexify $k_{\perp} = k_y $, and $ k_{\parallel} = k_x$ will be now one of the parameters determining the topological phase transition points. The two EP's for $A (k_x=k_{\parallel}, k_y=k_{\perp} ) =0$ are given by
\begin{equation}
k_{\perp}^{\pm} = -2 \, \ri \ln \Bigg [ \frac{- J_3 \,\exp (-\ri \, k_{\parallel}/2) \pm \sqrt{ J_3^2 \, \exp (- \ri\, k_{\parallel} )- 4\, J_1 J_2 } }
{2 \,J_2 }  \Bigg ] \,.
\label{armep}
\end{equation}
No Majorana zero mode exists for any value of $k_x$ if $ J_1 = J_2 = J_3 $ or $J_1 < J_2 $. For $ J_1 > J_2 $, a Majorana fermion can exist for a specific range of values for $k_x$. Figures~\ref{zigarm}~(c) and \ref{zigarm}~(d) show these topological phases, obtained using equation~(\ref{fchiral1}).

Equations~(\ref{zigzagep}) and (\ref{armep}) are seen to coincide with the solutions of the Majorana edge states obtained earlier by the transfer matrix formalism \cite{nakada1996,krishnendu}.

\section{EP formalism for the DIII class}
\label{d3}

A point defect in class DIII can support a Majorana Kramers pair (MKP) corresponding to doubly degenerate Majorana zero modes, whereas a line defect can support a pair of helical Majorana edge states. Both are characterized by a $\mathbb{Z}_2$ topological invariant. The chiral symmetry operator $\mathcal{O}$ can be defined such that the Hamiltonian in class DIII can be brought to the block off-diagonal form [equation~(\ref{chiralmom})], just like for the class BDI.

\subsection{1d model}
\label{d31d}

\begin{figure}[!t]
\centering
\begin{minipage}{0.49\linewidth}
\vspace{-0.7cm}
\begin{center}
\includegraphics[width=0.98\textwidth]{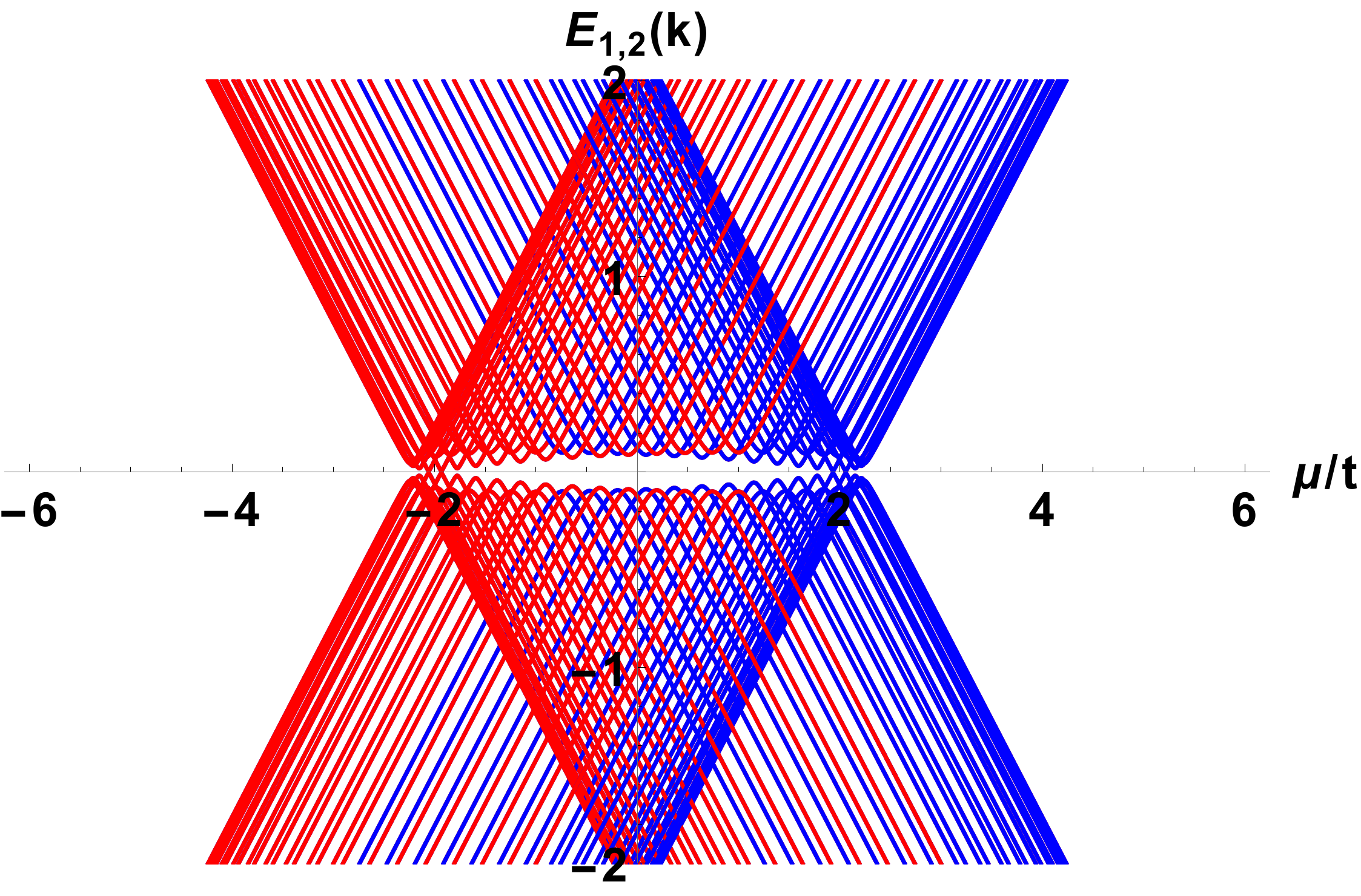}
\end{center}
\end{minipage}
\begin{minipage}{0.49\linewidth}
\begin{center}
\includegraphics[width=0.72\textwidth]{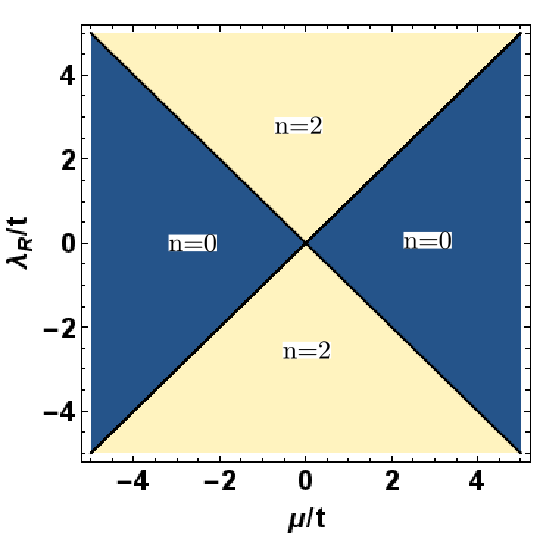}
\end{center}
\end{minipage}
\centerline{(a)~~~~~~~~~~~~~~~~~~~~~~~~~~~~~~~~~~~~~~~~~~~~~~~~~~~~~~~~~~~~~~~~~~~~~~~~~~~~~~~~~~~~~~~~~~~~~~~~~(b)}
\caption{\label{d3-1d}  (Color online)
Parameter: $ \Delta = 0.1 t $ corresponding to the Hamiltonian in equation~(\ref{perd31}).
(a)~Energy bands $E_{1,2} (k)$, given in equation~(\ref{epd31}), have been plotted in blue and red, respectively as functions of $\mu/t$, for $ \lambda_{\text{R}} = 2t $.
(b) $f(\mu/t, \lambda_{\text{R}}/t)$ giving the count ``$n$'' of the chiral Majorana fermions.
}
\end{figure}

A simple 1d model of topological superconductivity in the class DIII is described by the Hamiltonian \cite{zhang2013}
\begin{eqnarray}
\label{perd31}
H_{m1} =  \sum_{k} \Psi_k^{\dagger} \,
h_{m1} (k) \, \Psi_k \,,
\qquad  \Psi_{k} = (c_{k\uparrow},c_{k\downarrow},c_{-k\downarrow}^{\dagger},-c_{-k\uparrow}^{\dagger})^{\text{T}} ,  \qquad\qquad\qquad\nn
h_{m1} (k) =
\left[\,  \xi_{m1} (k) \, \sigma_0 + \lambda_{\text{R}} \sin(k) \, \sigma_z  \right ]  \tau_z
+ \,  \Delta \, \cos (k) \, \sigma_0 \tau_x \,,
\qquad  \xi_{m1}(k) =t \cos(k)-\mu \,.
\end{eqnarray}
This system may be realized in a Rashba wire that is proximity-coupled to a nodeless $s_{\pm}$ wave superconductor.
The energy eigenvalues are given by:
\begin{eqnarray}
E_1(k) = \pm
\sqrt {\left [ \,  \xi_{m1}(k) - \lambda_{\text{R}} \sin (k) \, \right ]^2
+ \Delta^2 \cos^2 (k) } \,,
\qquad E_2(k) = \pm
\sqrt {\left [ \, \xi_{m1}(k) + \lambda_{\text{R}} \sin (k) \, \right ]^2
+ \Delta^2 \cos^2 (k) } \,,
\label{epd31}
\end{eqnarray}
whose plots are shown in figure~\ref{d3-1d}~(a) for $\Delta = 0.1t$ and $ \lambda_{\text{R}} = 2t$, as $\mu/t$ is varied along the horizontal axis.

Observing that a chiral symmetry operator $\mathcal O = \sigma_0 \tau_y$ exists in the presence of $\mathcal{M}_z\,$, we rotate the Hamiltonian in equation~(\ref{perd31}) to the chiral basis, where it takes the form
\begin{eqnarray}
H_{m1}^{\text{chi}}(k) = U_4^{\dagger} \, h_{m1} (k) \, U_4
= \left(
\begin{array}{cc}
 0 & h_{m1}^u (k) \\
 h_{m1}^l (k)  &  0   \\
\end{array} \right) \,,\qquad
U_{4} =\frac{1}{\sqrt{2} }
\left( \begin{array}{cccc}
0 & -\ri & 0 & \ri \\
0 & 1 & 0 & 1 \\
-\ri & 0 & \ri & 0 \\
1 & 0 & 1 & 0 \\
\end{array} \right) \,,\nn
h_{m1}^u (k) =
\text{diag} \Big (
\ri \, \Delta  \cos (k) - \xi_{m1}(k)+\lambda_{\text{R}} \sin(k) ,\quad
\ri \, \Delta  \cos (k) - \xi_{m1}(k) - \lambda_{\text{R}} \sin(k) \Big )  \,,\quad\nn
h_{m1}^l (k) =
\text{diag} \Big (
- \ri\, \Delta  \cos (k) - \xi_{m1}(k)+\lambda_{\text{R}} \sin(k) ,\quad
- \ri \, \Delta  \cos (k) - \xi_{m1}(k) - \lambda_{\text{R}} \sin(k) \Big )  \,.
\label{hmchiral}
\end{eqnarray}

The solutions for the EP's are then given by either
\begin{eqnarray}
\det\left[h_{m1}^u  (k)\right]=0 \qquad\Rightarrow\qquad
\ri \, \Delta  \cos (k) - \xi_{m1}(k) = s_1  \lambda_{\text{R}} \sin(k) \nn
\Rightarrow\qquad k= k^{m 1 u}_{s_1,s_2}
= - \ri \ln \Bigg [ \frac
 {\mu + s_2
\sqrt{ \mu^2 -\lambda_{\text{R}}^2 -( t -\ri \, \Delta)^2  }
}
{ t - \ri \,(\Delta + s_1 \, \lambda_{\text{R}} )}
\Bigg ] \,, \quad\qquad
\label{detud31}
\end{eqnarray}
or
\begin{eqnarray}
\det\left[ h_{m1}^l (k)\right]=0 \qquad\Rightarrow\qquad
\ri\, \Delta  \cos (k) + \xi_{m1}(k)  = s_1  \lambda_{\text{R}} \sin(k) \nn
\Rightarrow \qquad k= k^{m 1l }_{s_1,s_2}
= - \ri \ln \Bigg [ \frac
 {\mu + s_2
\sqrt{ \mu^2 -\lambda_{\text{R}}^2 -( t + \ri \, \Delta)^2  }
}
{ t + \ri \,(\Delta + s_1 \, \lambda_{\text{R}} )}
\Bigg ] \,, \quad\qquad
\label{detld31}
\end{eqnarray}
where $ \left (s_1 =\pm 1 , s_2  = \pm 1 \right  ) $. For this model, we note that the two different sets of EP solutions, related by $ \lbrace \Im (k) \rbrace \big\vert_{ \text{set} = \text{A}} = - \lbrace \Im( k) \rbrace \big\vert_{ \text{set} = \text{B} } \,$, are obtained from equations~(\ref{detud31}) and (\ref{detld31}) when we set $s_1 = 1$ and $s_1 =-1$, respectively. This is related to the fact that $ [ h_{m1}^u  (k)]^{\dagger}\neq  [ h_{m1}^u (-k) ] ^{\text{T}} $ (where $ h_{m 1}^l (k)  =[ h_{m1}^u  (k)]^{\dagger} )$ for real $k$. However, we have argued before that for equation~(\ref{fchiral1}) to work, we must take all the EP solutions from one of the sets related by a negative sign of $\Im (k)$. Using either $( k^{m 1 u }_{+1,s_2}, k^{m 1 l }_{+1,s_2} )$ or $( k^{m1 u }_{-1,s_2}, k^{m 1l }_{-1,s_2} )$ (rather than both), figure~\ref{d3-1d}~(b) gives the correct topological phase diagram in the $\mu/t-\lambda_{\text{R}} /t$ plane, for $\Delta= 0.1t$. We clearly see that there exist phases with a pair of MBSs, which correspond to one Kramers doublet~(MKP).

\subsection{2d model}
\label{d32dd}

The 1d model of a Rashba semiconductor combined with a nodeless $s_{\pm}$ wave superconductor can be easily generalized to a 2d system, described by the Hamiltonian \cite{zhang2012}
\begin{eqnarray}
\label{perd32}
H_{m2} =  \sum_{ \textbf k} \Psi_{\textbf k}^{\dagger} \,
h_{m2} (\textbf k) \, \Psi_{ \textbf k} \,,
\qquad  \Psi_{ \textbf k} = (c_{{ \textbf k} \uparrow},c_{k\downarrow},c_{- { \textbf k}\downarrow}^{\dagger},-c_{-{ \textbf k} \uparrow}^{\dagger})^{\text{T}} , \qquad\qquad \qquad\nn
h_{m2} ({ \textbf k}) =
 \left[  \xi_{m 2} ({ \textbf k})   + \,  \Delta_m (\textbf k) \right]  \sigma_0 \tau_z
 + \, 2 \, \lambda_{\text{R}} [ \sin (k_x) \, \sigma_y -\sin (k_y) \,\sigma_x ] \tau_z
\, , \qquad\qquad\nn
\xi_{m 2}({ \textbf k}) = -2 t[ \cos(k_x) + \cos(k_y) ] -\mu \,,
\qquad \Delta_m (\textbf k) = \Delta_0 + 2 \, \Delta_1 [ \cos(k_x) + \cos(k_y)] \,,
\end{eqnarray}
where $ \Delta_m (\textbf k)$ is the $s_{\pm}$ wave singlet pairing potential that switches its sign between
the centre $ (0, 0)$ and the corner $ (\pi, \pi)$ of the 2d Brillouin zone, when $0 < |\Delta_0 | < 4 \Delta_1 $.

\begin{figure}[!t]
\begin{center}
\subfigure[]{\includegraphics[width=0.42\textwidth ]{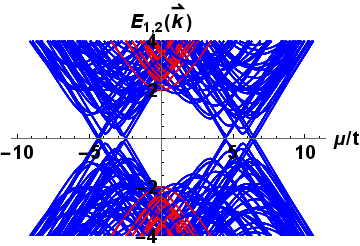}
\label{d32den}}\\
\subfigure[]{\includegraphics[width=0.35\textwidth ]{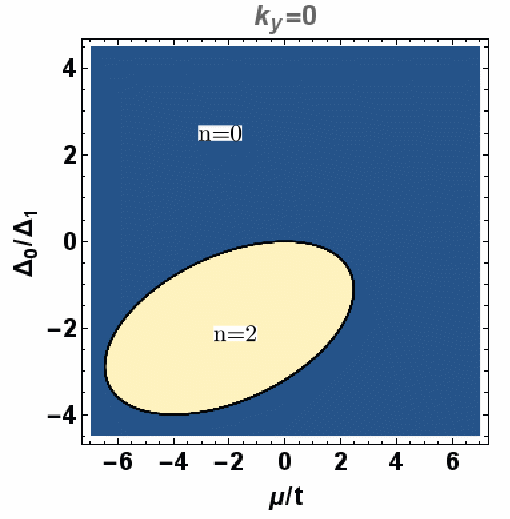}
\label{d32dphase1}} \qquad
\subfigure[]{\includegraphics[width=0.35\textwidth ]{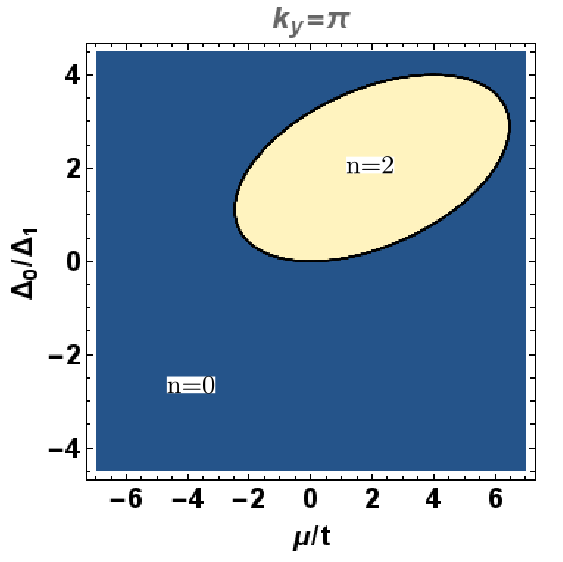}
\label{d32dphase2}}
\end{center}
\caption{\label{d32d}  (Color online) Panel (a) shows the plot of energy levels $E_{1,2}(\textbf k)$ of equation~(\ref{epd32d})  in blue and red, respectively, for the Hamiltonian in equation~(\ref{perd32}), as functions of $\mu / t$. We have used $\lambda_{\text{R}}=\Delta_0 =\Delta_1= 2t $. Panels (b) and (c) show the contourplots of $f(\mu/t,\,\Delta_0/\Delta_1,\, k_y,\, \lambda_{\text{R}}=2t,\, \Delta_1= t )$ in the $\Delta_0 /\Delta_1-\mu/t$ plane, giving the count ``$n$'' of the Majorana edge states along the $y$-direction, for $k_y = 0$ and $k_y=\pi$, respectively.}
\end{figure}

The energy eigenvalues are given by:
\begin{eqnarray}
&& E_1( \textbf k) =
\pm
\sqrt {\left [ \, \xi_{m 2}(\textbf k) -2\, \lambda_{\text{R}} \sqrt{ \sin^2(k_x) + \sin^2(k_y)} \, \right ]^2
+ \Delta_m^2 (\textbf k) }\,, \nn
&& E_2( \textbf k)
= \pm
\sqrt {\left [ \, \xi_{m 2}(\textbf k) + 2\, \lambda_{\text{R}} \sqrt{ \sin^2(k_x) + \sin^2(k_y)} \, \right ]^2
+ \Delta_m^2 (\textbf k) } \,,
\label{epd32d}
\end{eqnarray}
whose plots are shown in figure~\ref{d32d}~(a) for $\lambda_{\text{R}} = \Delta_0= \Delta_1= 2 t$, as $\mu/t$ is varied along the horizontal axis.

Rotating the Hamiltonian in equation~(\ref{perd32}) to the diagonal basis of the chiral symmetry operator $\mathcal O = \sigma_0 \tau_y\,$, we get
\begin{eqnarray}
H_{m2}^{\text{chi}}( \textbf k) = U_4^{\dagger} \, h_{m 2} (\textbf k) \, U_4
= \left(
\begin{array}{cc}
 0 & h_{m 2}^u (\textbf k) \\
 h_{m 2}^l (\textbf k)  &  0   \\
\end{array} \right) \,, \qquad\qquad\nn
h_{m2}^u ( \textbf k)
=
\left(
\begin{array}{cc}
\ri \Delta_m (\textbf k) -\xi_{m2} (\textbf k)
& 2\, \lambda_{\text{R}} [   \sin(k_y) - \ri \sin(k_x) ]   \\
 2\, \lambda_{\text{R}} [   \sin(k_y) + \ri \sin(k_x) ]
 &  \ri \Delta_m (\textbf k) -\xi_{m2} (\textbf k)  \\
\end{array} \right)
,\nn
h_{m 2}^l ( \textbf k)
=
\left(
\begin{array}{cc}
-\ri \Delta_m (\textbf k) -\xi_{m2} (\textbf k)
& 2\, \lambda_{\text{R}} [  \sin(k_y) - \ri \sin(k_x)  ]   \\
 2\, \lambda_{\text{R}} [   \sin(k_y) + \ri \sin(k_x)  ]
 &  - \ri \Delta_m (\textbf k) -\xi_{m2} (\textbf k)  \\
\end{array} \right)
 .
\label{hm2chiral}
\end{eqnarray}
The equations for the EP's, corresponding to edge modes along the $y$-direction (so that $k_{\parallel}= k_y$ and $k_{\perp} = k_x $), for $k_y =0$ and $k_y=\pi$, are given by
\begin{eqnarray}
\det \left [h_{m 2}^u  (\textbf k) \right ] \Big\rvert_{k_y=(0,\pi)}=0
\qquad  \Rightarrow\qquad
\lbrace \ri \Delta_m (\textbf k) - \xi_{m 2}(\textbf k) \rbrace \big\rvert_{k_y=(0,\pi)}= 2 \, s_1  \lambda_{\text{R}} \sin(k_x) \,,
\label{detud321}
\end{eqnarray}
and
\begin{eqnarray}
\det \left [h_{m 2}^l  (\textbf k) \right ] \Big\rvert_{k_y=(0,\pi)}=0
\qquad \Rightarrow\qquad
\lbrace \ri \Delta_m (\textbf k) + \xi_{m 2}(\textbf k) \rbrace \big\rvert_{k_y=(0,\pi)}= 2\, s_1  \lambda_{\text{R}} \sin(k_x) \,.
\label{detld322}
\end{eqnarray}
The solutions for $k_y =0 $ are:
\begin{eqnarray}
k^{m 2 u}_{s_1,s_2}
=
 - \ri \ln
 \Bigg [ - \frac
 { 2t +\mu + \ri\, (2 \, \Delta_1 + \Delta_0 )
  }
{ 2t + 2 \, \ri \,(\Delta_1 + s_1 \, \lambda_{\text{R}} )}
+ \, \frac  {  s_2
\sqrt{
( \ri\, \mu- \Delta_0)
\, (4 \, \Delta_1 + \Delta_0  - \ri\, \mu - 4 \, \ri \, t) -4 \, \lambda_{\text{R}}^2
  }
}
{ 2 t +  2 \, \ri \,(\Delta_1 + s_1 \, \lambda_{\text{R}} )}
\Bigg ] \,,
\label{ud301}
\end{eqnarray}
and
\begin{eqnarray}
 k^{m 2l }_{s_1,s_2}
 =
 - \ri \ln
 \Bigg [  \frac
 { -2 t - \mu + \ri\, (2 \, \Delta_1 + \Delta_0 )
  }
{ 2 t -  2 \, \ri \,(\Delta_1 + s_1 \, \lambda_{\text{R}} )}
+ \, \frac  {  s_2
\sqrt{
- ( \ri\, \mu +\Delta_0)
\, (4 \, \Delta_1 + \Delta_0  + \ri\, \mu + 4 \, \ri \, t) -4 \, \lambda_{\text{R}}^2
  }
}
{ 2 t  -   2 \, \ri \,(\Delta_1 + s_1 \, \lambda_{\text{R}} )}
\Bigg ] \,.
\label{ud302}
\end{eqnarray}
Those for $k_y =\pi $ are:
\begin{eqnarray}
k^{m 2 u}_{s_1,s_2} \
=
 - \ri \ln \Bigg [ \frac
 { 2t -\mu + \ri\, (2 \, \Delta_1-\Delta_0 )
  }
{ 2 t +  2 \, \ri \,(\Delta_1 + s_1 \, \lambda_{\text{R}} )}
+ \, \frac  {  s_2
\sqrt{
(\Delta_0 - \ri\, \mu)
\, (4 \, \Delta_1 -\Delta_0  + \ri\, \mu - 4 \, \ri \, t) -4 \, \lambda_{\text{R}}^2
  }
}
{ 2 t +  2 \, \ri \,(\Delta_1 + s_1 \, \lambda_{\text{R}} )}
\Bigg ] \,,
\label{ud3pi1}
\end{eqnarray}
and
\begin{eqnarray}
 k^{m 2l }_{s_1,s_2} =
 - \ri \ln \Bigg [ \frac
 { 2 t -\mu + \ri\, (\Delta_0 - 2 \, \Delta_1)
  }
{ 2 t -  2 \, \ri \,(\Delta_1 + s_1 \, \lambda_{\text{R}} )}
+ \, \frac  {  s_2
\sqrt{
(\Delta_0 + \ri\, \mu)
\, (4 \, \Delta_1 -\Delta_0  - \ri\, \mu + 4 \, \ri \, t) -4 \, \lambda_{\text{R}}^2
  }
}
{ 2  t -  2 \, \ri \,(\Delta_1 + s_1 \, \lambda_{\text{R}} )}
\Bigg ] \,.
\label{ud3pi2}
\end{eqnarray}
Here, $k^{m 2 u}_{s_1,s_2}$ and $k^{m 2 l}_{s_1,s_2}$ correspond to the vanishing of $\det [h_{m 2}^u  (\textbf k) ]$ and $\det[h_{m 2}^l  (\textbf k) ]$, respectively, and $ \left (s_1 =\pm 1 , s_2  = \pm 1 \right  ) $. Two distinct sets of EP solutions, related by $ \lbrace \Im (k_{\perp} ) \rbrace \big\vert_{ \text{set} = \text{A}} = - \lbrace \Im( k_{\perp}) \rbrace \big\vert_{ \text{set} = \text{B}} \,$, are obtained by setting $s_1 = 1$ and $s_1 =-1$, respectively. Using either $( k^{m 2 u }_{+1,s_2}, k^{m 1 l }_{+1,s_2} )$ or $( k^{m 2 u }_{-1,s_2}, k^{m 1l }_{-1,s_2} )$ (rather than both) in equation~(\ref{fchiral1}), figures~\ref{d32d}~(b) and \ref{d32d}~(c) give the desired topological phase diagrams in the $\mu/t-\Delta_0 / \Delta_1$ plane, for $\lambda_{\text{R}} = 2t$ and $\Delta_1 = t $. The topological phases with a pair of helical Majorana edge states are clearly seen.

\section{Broken time reversal symmetry: class D}
\label{classd}

In this section, we consider 1d and 2d Hamiltonians in the symmetry class D, where the TRS is broken. We will see that the EP formalism in the complex $k_{\perp}$-plane is not applicable for such systems.


\subsection{1d spinless model}
\label{classd1d}

We examine the spinless model described by the Hamiltonian
\begin{eqnarray}
\label{hd1}
H_{D1}   =
\sum_{j=1}^{N-1 }  \left( - w \, c_j^{\dagger}\, c_{j+1} + \Delta \, c_j \, c_{j+1} - w^* \, c_{j+1}^{\dagger}\, c_{j} + \Delta^* \, c_{j+1} \, c_{j} \right)
- \sum_{j=1}^{N} \mu \left(  c_j^{\dagger}\, c_j - \frac{1}{2} \right) \,,
\end{eqnarray}
which looks similar to $H_K$ in equation~(\ref{kitaev-open}), but with the important difference that the TRS is broken by the fact that $w$ and $\Delta$ can be complex numbers \cite{diptiman2}. Without any loss of generality, we can choose $\Delta$ to be real and encode the entire phase-difference ($\phi$) between $\Delta$ and $w_0$ by writing $w=w_0 \, \re^{\ri \phi}$, where $w_0$ is real and positive.

With PBC's, one can write the corresponding BdG Hamiltonian in the momentum space as:
\begin{eqnarray}
\label{perd1}
H_{D1} = - \sum_{ k }
\left(
\begin{array}{cc}
 c_k^{\dagger}  & c_{-k}
\end{array} \right)
\, h_{D1} (k)
\left(
\begin{array}{c}
 c_k  \\
 c_{-k}^{\dagger}
\end{array} \right)\,,\qquad\qquad\qquad\qquad\quad \nn
h_{D1} (k) =
\left[ 2 \, w_0  \sin ( \phi ) \sin ( k ) \right]  \tau_0
- \left[ 2 \, w_0  \cos( \phi ) \cos ( k ) + \mu \right]  \tau_z
+ \, 2\, \Delta \sin \left( k \right)  \tau_y \,.
\end{eqnarray}
The energy eigenvalues are given by:
\begin{eqnarray}
 E(k) = 2 \, w_0 \, \sin ( \phi ) \,\sin ( k )
\pm \tilde e_t \,,
\qquad \tilde e_t = \sqrt{
\left[ 2 \, w_0 \, \cos ( \phi ) \,\cos ( k ) + \mu \right]^2 + 4\, \Delta^2  \, \sin^2(k)
}\,.
\end{eqnarray}
Hence, it follows that two levels become degenerate when
\begin{eqnarray}
\tilde e_t = 0 \qquad
\Rightarrow\qquad 2 \, w_0 \, \cos ( \phi ) \,\cos ( k ) + \mu
= \pm \, 2 \, \ri \, \Delta \sin (k)  \,.
\label{epd1}
\end{eqnarray}

\begin{figure}[!b]
\begin{center}
\subfigure[]{\includegraphics[width=0.3\textwidth]{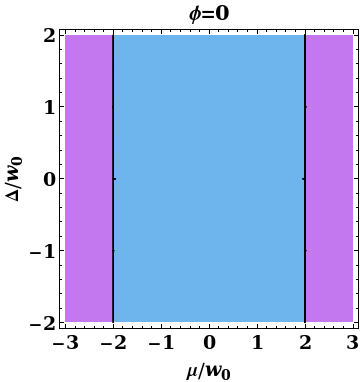}
\label{dd1}} \qquad
\subfigure[]{\includegraphics[width=0.3\textwidth]{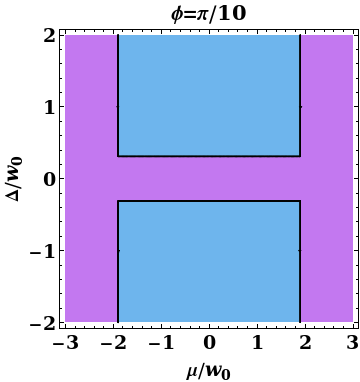}
\label{dd2}}\\
\subfigure[]{\includegraphics[width=0.3 \textwidth]{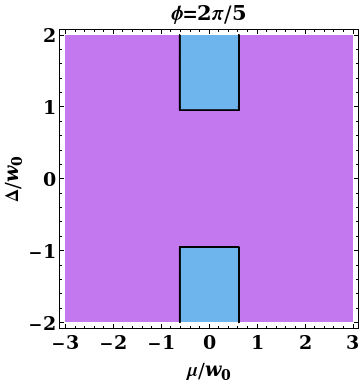}
\label{dd3}} \qquad
\subfigure[]{\includegraphics[width=0.3\textwidth ]{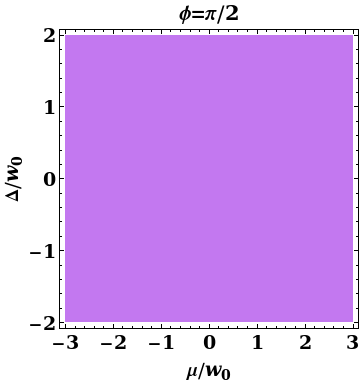}
\label{dd4}}
\end{center}
\caption{\label{dd}  (Color online) Panels (a), (b), (c) and (d) show the contourplots of $f(\mu,\Delta)$ corresponding to the Hamiltonian in equation~(\ref{perd1}), giving the count ``$n$'' of the MBSs in the  ${\mu}/ w_0-\Delta/w_0$ plane for $\phi =0, \, \pi/10, \, 2 \pi/5 $ and $ \pi/2$, respectively. The blue regions have $n=1$ MBS at each end of the open chain, while the purple regions correspond to the $n=0$ trivial phases.
}
\end{figure}

When extended to the complex $k$-space, there can be EP's where the norm of an eigenvector vanishes. For convenience, we rotate $h_{D1}(k)$ to find the points (EP's) where the Hamiltonian becomes non-diagonalizable:
\begin{eqnarray}
h_{D1,\text{od}} (k) = U_{D1}^{\dagger} \, h_{D1}(k) \, U_{D1}
= \left(
\begin{array}{cc}
 2 \, w_0 \, \sin ( \phi ) \,\sin ( k ) & A_{D1}(k) \\
 B_{D1} (k)  & 2 \, w_0 \, \sin ( \phi ) \,\sin ( k )  \\
\end{array} \right) \,,\nn
U_{D1} = \frac{\ri}{\sqrt{2}}
\left( \begin{array}{cc}
 1 & -1 \\
 -1   &  -1   \\
\end{array} \right) \,,\qquad\qquad\qquad\qquad\qquad\qquad\nn
A_{D1} (k) =  2 \, w_0 \, \cos ( \phi ) \,\cos ( k ) + \mu
+ 2 \, \ri \, \Delta \sin (k) \,,\qquad\qquad\qquad\nn
B_{D1} (k) = 2 \, w_0 \, \cos ( \phi ) \,\cos ( k ) + \mu
- 2 \, \ri \, \Delta \sin (k) \,. \qquad\qquad\qquad
\end{eqnarray}
The EP's are given by $A_{D1} (k)=0$ or $B_{D1}(k)=0$. At such points, two levels coalesce for a complex value of $k$ satisfying equation~(\ref{epd1}).

However, we immediately observe that these EP's do not correspond to zero energy modes, which appear when
\begin{eqnarray}
\det \left[ h_{D1} (k) \right] = 0 \qquad
\Rightarrow\qquad \left[ 2 \, w_0 \, \cos ( \phi ) \,\cos ( k ) + \mu \right]^2 + 4\, \Delta^2   \sin^2(k)
 = 4\,  w_0^2 \, \sin ^2 ( \phi ) \,\sin ^2 ( k )
\end{eqnarray}
is satisfied. We note that since $\det [ h_{D1} (k) ]$ is equal to the product of the energy eigenvalues, vanishing of $\det [ h_{D1} (k) ]$ implies the condition for the existence of a zero energy solution.

Although the EP description no longer applies now to the existence of MBSs, we can find the complex $k$-values satisfying
\begin{eqnarray}
\det \left[ h_{D1,\text{od}} (k)  \right] =0 \qquad
\Rightarrow\qquad
 2 \, w_0 \, \cos ( \phi ) \,\cos ( k ) + \mu
 =  \pm \, 2 \, \ri \sin (k) \, \sqrt{\Delta^2 - w_0^2  \sin ^2 ( \phi ) } \,.
\end{eqnarray}
We can solve for the $k$-values either with the ``$+$'' or the ``$-$'' sign on the RHS (rather than both), and plug in the roots of that equation into the formula in equation~(\ref{fchiral1}). The function $f (\mu,\Delta)$ will still give the number of MBSs in a given topological phase. Once again we emphasize that to count the zero modes in equation~(\ref{fchiral1}), we should include only one of the two sets of roots related by a sign change of $\Im (k) $, as these two sets correspond to the wavefunctions of the pair of MBSs at the two opposite edges. We have shown the plots of $f (\mu,\Delta)$ in figure~\ref{dd} for four different values of $\phi$. It indeed captures the correct $\mathbb{Z}_2$ topological invariant ($n=0,$ or $1$). No zero mode exists, i.e., the system is entirely gapped in certain regions in the $\phi-\mu$ plane where $\Im \left(k \right )$ vanishes, as $\Im  \left(k \right )$ is related to the exponential part of the MBS wavefunction in the real space.


\subsection{2d spinless model}
\label{classd2d}



\begin{figure}[!t]
\vspace{-1mm}
\begin{center}
\subfigure[]{\includegraphics[width=0.42\textwidth]{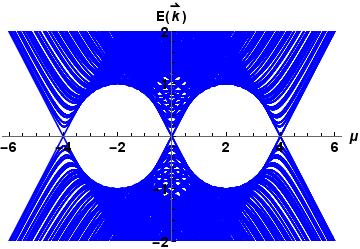}
\label{d2den}}\\
\subfigure[]{\includegraphics[width=0.35\textwidth]{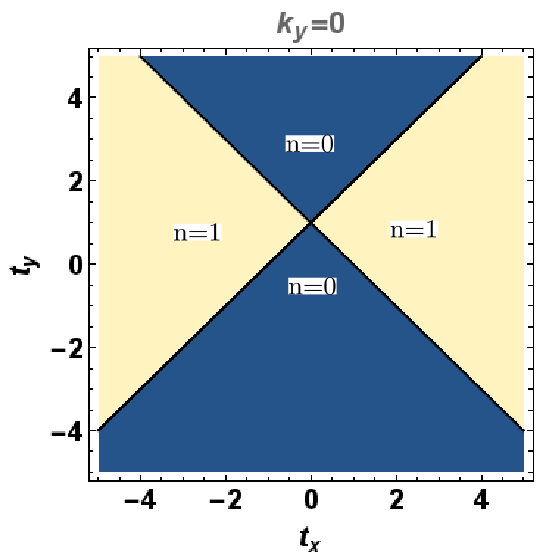}
\label{d2dphase1}} \qquad
\subfigure[]{\includegraphics[width=0.35\textwidth]{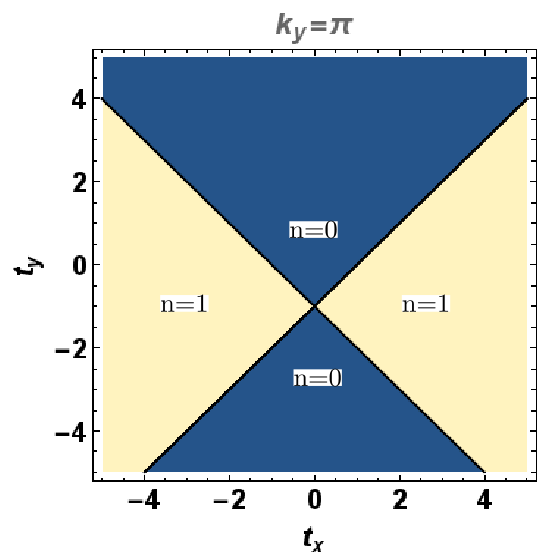}
\label{d2dphase2}}
\end{center}
\vspace{-3mm}
\caption{\label{d2d}  (Color online) Parameters: $ d_x=d_y=1  $ corresponding to the Hamiltonian in equation~(\ref{perd2d}).
Panel (a) shows the plot of energy levels $E (k)$, given in equation~(\ref{epd2d}), as functions of $\mu$, for $t_x=t_y= 1$. Panels (b) and (c) show the contourplots of $f(t_x,t_y, k_y, \mu=2)$ in the $t_x-t_y$ plane, giving the count ``$n$'' of the non-chiral Majorana edge states along the $y$-direction, for $k_y = 0$ and $k_y=\pi$, respectively.}
\end{figure}

The following Hamiltonian gives a model of a $p + \ri p$ wave superconductor on a square lattice \cite{nagaosa2012}:
\begin{eqnarray}
\label{perd2d}
H_{D2} =- \sum_{\textbf k }
\left(
\begin{array}{cc}
 c_{\textbf k}^{\dagger}  & c_{-\textbf k}
\end{array} \right)
\, h_{D2} (\textbf k)
\left(
\begin{array}{c}
 c_{\textbf k}  \\
 c_{- \textbf k}^{\dagger}
\end{array} \right)\,, \qquad\qquad\qquad\qquad\nn
h_{D2} (\textbf k) =
\left[ 2 \, t_x  \cos ( k_x ) + 2\, t_y \cos ( k_y )- \mu \right]  \tau_z
+ \, d_x \sin (k_x) \, \tau_x + d_y \sin (k_y) \, \tau_y \,,
\end{eqnarray}
where $(t_x\,, t_y)$ are the hopping strengths and $(d_x \,, d_y)$ are the pairing amplitudes along the $(x,y)$-direc\-tions, and $\mu$ is the chemical potential.
The energy eigenvalues, given by
\begin{eqnarray}
E(\textbf k) = \pm \sqrt{
\left[ 2 \, t_x  \cos ( k_x ) + 2\, t_y \cos ( k_y )- \mu \right]^2
+ \, d_x^2 \sin^2 (k_x)  + d_y^2 \sin^2 (k_y)
}\,,
\label{epd2d}
\end{eqnarray}
are plotted in figure~\ref{d2d}~(a) for $t_x=t_y=d_x = d_y=1$, as $\mu$ is varied along the horizontal axis.

We consider the edges parallel to the $y$-axis, so that $k_{\parallel} = k_y$ and $ k_{\perp} = k_x\,$. Complexifying $k_x\,$, we can compute the number of non-chiral Majorana fermions propagating along these edges with momenta $k_y =0$ and $ k_y = \pi$, using equation~(\ref{fchiral1}). The complex $k_x\,$-values satisfying $ \det[ h_{D2} (\textbf k) ] =0$, for $k_y=0$ and $k_y = \pi$, are given by
\begin{eqnarray}
2\, t_x \cos (k_x) + 2\, t_y -\mu
= \ri \, s_1\,  d_x \sin (k_x)\qquad\qquad\qquad\qquad\qquad\qquad\qquad
\nn \Rightarrow\qquad k_x = k_{s_1,s_2}
= -\ri \, \ln \Bigg [
 \frac{2\, s_1 (1-t_y) + s_2 \sqrt{4 (t_y-1)^2 + d_x^2 -4\, t_x^2 }}
 {d_x + 2\, s_1 \, t_x}
 \Bigg ]\,,
 \end{eqnarray}
 and
\begin{eqnarray}
2\, t_x \cos (k_x) - 2\, t_y -\mu
= \ri \, s_1\,  d_x \sin (k_x)\qquad\qquad\qquad\qquad\qquad\qquad\qquad
\nn \Rightarrow\qquad k_x = k_{s_1,s_2}
 = -\ri \, \ln \Bigg [
 \frac{2\, s_1 (1 + t_y) + s_2 \sqrt{4 (t_y + 1)^2 + d_x^2 -4\, t_x^2 }}
 {d_x + 2\, s_1 \, t_x}
 \Bigg ]\,,
\end{eqnarray}
respectively. Here, $ \left (s_1 =\pm 1 , s_2  = \pm 1 \right  ) $, and we need to use either $s_1=1 $ or $s_1=-1$ (rather than both) in equation~(\ref{fchiral1}), to obtain the phase diagrams shown in figures~\ref{d2d}~(b) and \ref{d2d}~(c).

\section{Conclusion}
\label{conclusion}

We have established the relation of the EP solutions for complexified momenta to the Majorana fermion wavefunctions bound to a topological defect in a system with a chiral symmetry, by studying some explicit examples in 1d and 2d. These models include both spinless and spinful cases. We have shown that such EP solutions cannot exist for systems in class D, where there is no chiral symmetry. The generic formula, which was proposed earlier \cite{ipsita-proof} to count the number of Majorana zero modes in arbitrary dimensions, has been demonstrated to chart out the desired topological phase diagrams for the wide variety of systems we have considered. The detailed study of these models also helps us illustrate how one distinct set of complex $k_{\perp}$-solutions for $\det [H_{\text{{BdG}}} (\textbf{k}) ] =0 $, related by $ \lbrace  \Im (k_{\perp} ) \rbrace \big\vert_{ \text{set} = \text{A} } = - \lbrace   \Im (k_{\perp} ) \rbrace \big\vert_{ \text{set} = \text{B}} \,$, should be used while using our formula. An explicit proof of the counting formula has also been discussed. For a system with or without a chiral symmetry, the imaginary parts of these solutions in the complexified $k_{\perp}$-plane are related to the exponential decay of the Majorana fermion wavefunctions in the bulk in the position space. Hence, the imaginary parts of $ \tilde n$ of the solutions in one set undergoes a change of sign across a topological phase transition point, if the number of Majorana zero modes at a defect changes by $\tilde n$.


\section{Acknowledgements}
We thank Atri Bhattacharya, Fiona Burnell, Sudip Chakravarty, Sumathi Rao, Diptiman Sen, Krishnendu Sengupta and Sumanta Tewari for stimulating discussions. We are also grateful
to Chen-Hsuan Hsu  and Arijit Saha for their valuable comments on the
manuscript. This research was partially supported by the Templeton Foundation.
Research at the Perimeter Institute is supported
in part by the Government of Canada
through Industry Canada, and by the Province of Ontario through the
Ministry of Research and Information.

\vspace{-2mm}
\appendix
\section{Choice of EP solutions}
\label{append:eg}

\begin{figure}[!t]
\vspace{-2mm}
\begin{center}
\subfigure[]{\includegraphics[width=0.45\textwidth]{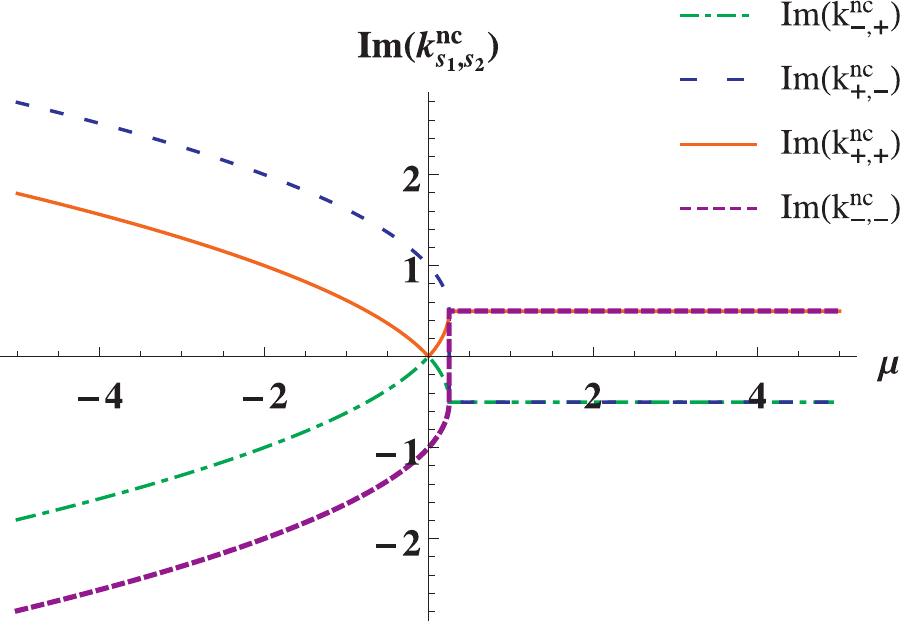}
\label{pnc0}} \qquad
\subfigure[]{\includegraphics[width=0.45\textwidth]{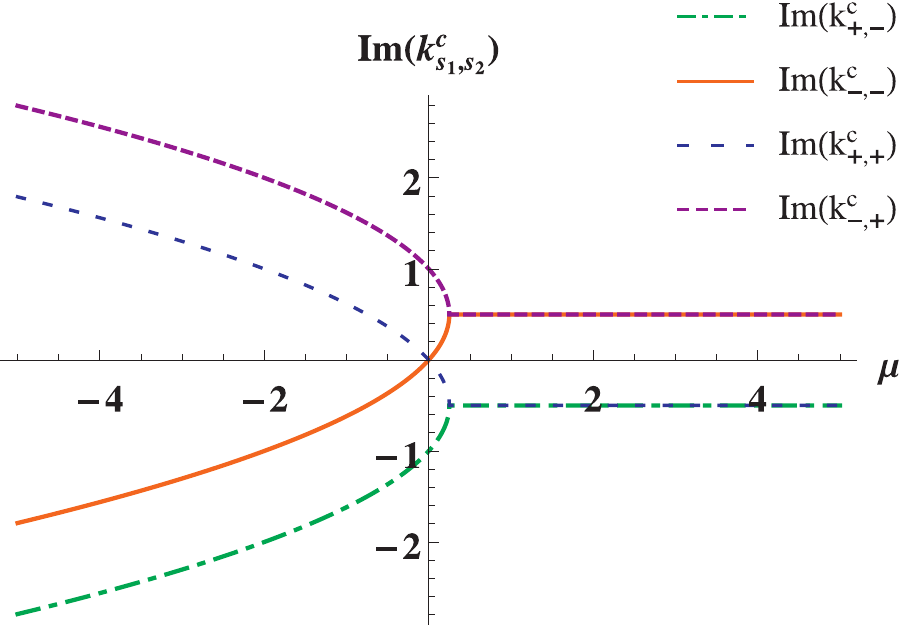}
\label{pc0}}\\
\vspace{-2mm}
\subfigure[]{\includegraphics[width=0.35\textwidth]{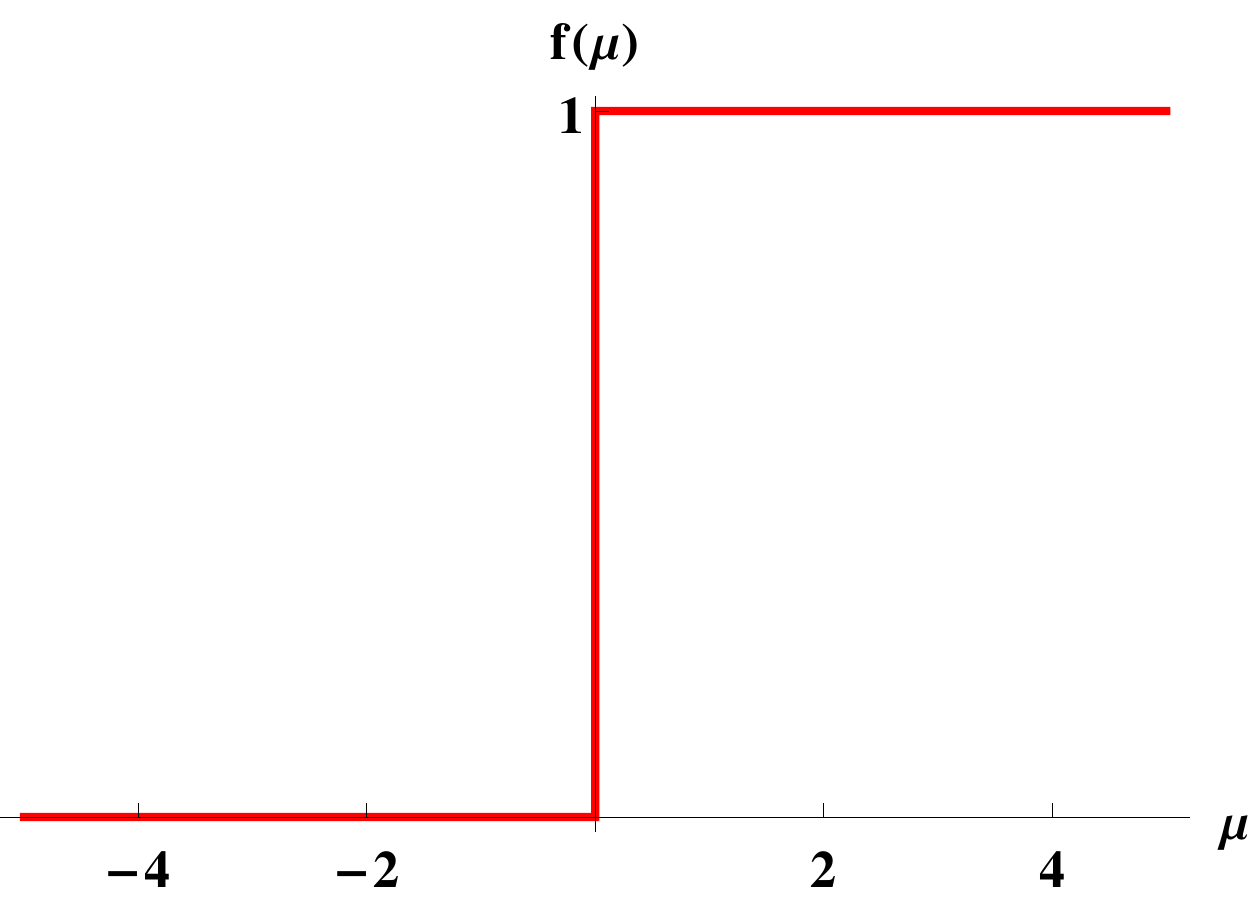}
\label{fnread}}
\end{center}
\vspace{-3mm}
\caption{\label{kcnc}  (Color online)
For edges of the system described by equation~(\ref{eg}), corresponding to $k_y =0 $:
(a) Plots of  $ \Im ( k^{nc}_{s_1,s_2}  )  $ versus $\mu$.
(b) Plots of  $ \Im  ( k^{c}_{s_1,s_2} )  $ versus $\mu$.
(c) $f(\mu)$ giving the count of the Majorana zero modes as a function of $\mu$.
}
\end{figure}

In this appendix, we provide a simple example to show how one should choose the correct EP solutions such that their imaginary parts are continuous functions in the parameter space for our counting formula\footnote{We thank Victor Gurarie for suggesting to clarify this point.}. Let us take the 2d class D Hamiltonian \cite{read}:
\begin{eqnarray}
\label{eg}
H_{D 3 } =   \sum_{\textbf k }
\left(
\begin{array}{cc}
 c_{\textbf k}^{\dagger}  & c_{-\textbf k}
\end{array} \right)
\, h_{D  3} (\textbf k)
\left(
\begin{array}{c}
 c_{\textbf k}  \\
 c_{- \textbf k}^{\dagger}
\end{array} \right)\,, \qquad
h_{D 3} (\textbf k) =
k_x \, \tau_x  +  k_y \, \tau_y + (\textbf k^2-\mu) \, \tau_z \,,
\end{eqnarray}
where $\mu$ is the chemical potential. The system is known to be topological for $\mu>0 $ and non-topological for $\mu<0 $. For non-chiral Majorana fermions along the edges parallel to the $y$-axis with momentum $k_y =0$, one should solve for $ \det [ h_{D 3 } ( k_x, k_y=0 ) ] = - k_x^2 - (k_x - \mu)^2    = 0$. We will have four solutions which can be written as either
\begin{eqnarray}
\label{inel}
 k_{s_1,s_2} ^{nc} =s_1 \sqrt{
 -\frac{1} {2}  + s_2  \frac{ \sqrt{1- 4 \, \mu} }  { 2 } +\mu
 }
 \,;
 \end{eqnarray}
 or
 \begin{eqnarray}
 \label{el}
\mathcal{F}_{s_1 } \equiv
 \left(    k_x^2 -\mu \right) +  \ri \, s_1  k_x =0
\qquad \Rightarrow\qquad k_x = k_{s_1,s_2}^{c}
=  \frac {   - \ri \, s_1 + s_2 \sqrt{4 \mu -1 } }
 { 2 }
\,,
 \end{eqnarray}
 where
\begin{eqnarray}
\det \big[ h_{D 3 } ( k_x, k_y=0 ) \big]  \equiv - \mathcal{F}_+  \mathcal{F}_- \,,
\end{eqnarray}
and $ \left (s_1 =\pm 1 , s_2  = \pm 1 \right  ) $.

The plots of $\Im ( k^{nc}_{s_1,s_2} )$, $\Im ( k^{c}_{s_1,s_2} )$ and $f(\mu)$ have been shown in figure~\ref{kcnc}. We find that $  \Im [ k^{nc}_{ +, -}  (\mu ) ]= -   \Im [ k^{nc}_{-,-}  (\mu ) ] $
and $  \Im [ k^{nc}_{ +, + }  (\mu ) ]= -   \Im [ k^{nc}_{-,+}  (\mu ) ] $.
Also, $  \Im [ k^{c}_{ +, + } (\mu ) ]= -   \Im [ k^{c}_{-,-}(\mu ) ] $
and $  \Im [ k^{c}_{ +, - }  (\mu ) ]=$ \linebreak$ -   \Im [ k^{c}_{-, + }  (\mu ) ] $.
So naively, one might think that from the solution set in equation~(\ref{inel}), we can evaluate $f (\mu)$ by using any one of the four pairs given by $ (  k^{nc}_{ +, -}, k^{nc}_{ +, + }  )$, $ (  k^{nc}_{ +, -}, k^{nc}_{ -, + }  )$, $ (  k^{nc}_{ -, -}, k^{nc}_{ -, + }  )$, $ (  k^{nc}_{ - , -}, k^{nc}_{ +, + }  )$.
Whereas, from the solution set in equation~(\ref{el}), $f (\mu)$ is expected to be obtained by using any one of the four pairs given by $ (  k^{ c }_{ +, + }, k^{ c}_{ +, - }  )$, $ (  k^{ c}_{ +, + }, k^{ c}_{ -, + }  )$, $ (  k^{ c}_{ -, -}, k^{ c}_{ +, - }  )$, $ (  k^{ c}_{ - , -}, k^{ c}_{ -, + }  )$.
One can check that this is true for the set in equation~(\ref{el}), rather than for the one in equation~(\ref{inel}). In other words, one gets the wrong phase diagram on using the first set. This is because all the $\Im (  k_{s_1,s_2} ^{nc} )$'s are not continuous functions of $\mu $.


\ukrainianpart

\title[]{Підрахунок  зв'язаних станів Майорани з використанням комплексних імпульсів}

\author[]{I. Мандал}
\address{Інститу теоретичної фізики ``Периметр'', Ватерлоо, Онтаріо N2L 2Y5, Канада}

\makeukrtitle

\begin{abstract}
Нещодавно (EPL, 2015, \textbf{110}, 67005) було встановлено зв’язок мiж
 фермiонами Майорани, зв’язаними з дефектами у довiльнiй вимiрностi, i
 комплексними iмпульсними коренями детермiнанта вiдповiдного об'ємного
 гамiльтонiану Боголюбова-де Жена. Базуючись на цьому розумiннi, запропоновано
 формулу для пiдрахунку числа ($n$) зв'язаних станiв Майорани з нульовою
 енергiєю, якi пов'язанi з топологiчною фазою системи. В цiй статтi дається
 вивiд формули пiдрахунку, яка застосовується до низки 1d i 2d моделей, що
 належать до класiв BDI, DIII i D. Показано, як можна успiшно побудувати
 топологiчнi фазовi дiаграми. Вивчення даних прикладiв дозволяє явно
 спостерiгати вiдповiднiсть мiж цими комплексними розв’язками для iмпульсу в
 Фур'є просторi i локалiзованими хвильовими функцiями фермiонiв Майорани в
 позицiйному просторi. Накiнець, пiдтверджено факт, що для систем з хiральною
 симетрiєю цi розв’язки є так званими ``винятковими точками'', де два чи бiльше
 власних значень ускладненого гамiльтонiана зливаються.

\keywords     виняткові точки, ферміони Майорани, BDI, DIII, D, підрахунок
\end{abstract}

\end{document}